\documentclass[lettersize,journal]{IEEEtran}
\usepackage{amsmath,amssymb,amsfonts}
\usepackage{algorithm}
\usepackage{algorithmic}
\usepackage{array}
\usepackage[caption=false,font=normalsize,labelfont=sf,textfont=sf]{subfig}
\usepackage{textcomp}
\usepackage{stfloats}
\usepackage{url}
\usepackage{bm}
\usepackage{verbatim}
\usepackage{graphicx}
\usepackage{cite}
\usepackage{booktabs}
\usepackage{cite}
\usepackage{xcolor}
\begin{document}

\newcounter{proposition}
\setcounter{proposition}{0}
\newenvironment{proposition}[1][]{\refstepcounter{proposition}\textbf{Proposition \arabic{proposition}:} \rmfamily}{\medskip}

\title{Hybrid Near-field and Far-field Localization with Holographic MIMO}

\author{
		Mengyuan~Cao,~\IEEEmembership{Student Member, IEEE},
        Haobo~Zhang,~\IEEEmembership{Member, IEEE},
		Yonina C. Eldar,~\IEEEmembership{Fellow, IEEE},
        Hongliang~Zhang,~\IEEEmembership{Member, IEEE}.
        % <-this % stops a space
		\thanks{
			M. Cao and Hongliang Zhang are with State Key Laboratory of Photonics and Communications, School of Electronics, Peking University, Beijing 100871, China (e-mail: \{caomengyuan, hongliang.zhang\}@pku.edu.cn). 

            Haobo Zhang is with the School of Electronic and Computer Engineering, Peking University Shenzhen Graduate School, Shenzhen 518055, China (e-mail: haobo.zhang@pku.edu.cn).
            
            Y. C. Eldar is with the Department of Mathematics and Computer Science, Weizmann Institute of Science, Rehovot 7610001, Israel (e-mail: yonina.eldar@weizmann.ac.il). 
            
			Part of this work has been published in IEEE WCNC 2024~\cite{b62}.
			}
}



\maketitle

\begin{abstract}
	Due to its ability to precisely control wireless beams, holographic multiple-input multiple-output (HMIMO) is expected to be a promising solution to achieve high-accuracy localization. However, as the scale of HMIMO increases to improve beam control capability, the corresponding near-field (NF) region expands, indicating that users may exist in both NF and far-field~(FF) regions with different electromagnetic transmission characteristics. As a result, existing methods for pure NF or FF localization are no longer applicable. We consider a hybrid NF and FF localization scenario in this paper, where a base station~(BS) locates multiple users in both NF and FF regions with the aid of a reconfigurable intelligent surface~(RIS), which is a low-cost implementation of HMIMO. In such a scenario, it is difficult to locate the users and optimize the RIS phase shifts because whether the location of the user is in the NF or FF region is unknown, and the channels of different users are coupled. To tackle this challenge, we propose a RIS-enabled localization method that searches the users in both NF and FF regions and tackles the coupling issue by jointly estimating all user locations. We derive the localization error bound by considering the channel coupling and propose an RIS phase shift optimization algorithm that minimizes the derived bound. Simulations show the effectiveness of the proposed method and demonstrate the performance gain compared to pure NF and FF techniques.
\end{abstract}

\begin{IEEEkeywords}
Holographic multiple-input multiple-output~(HMIMO), reconfigurable intelligent  surface~(RIS), near-field localization, far-field localization.
\end{IEEEkeywords}

\section{Introduction}

Future sixth-generation (6G) wireless systems will support a wide range of applications such as autonomous driving and augmented reality\cite{b46} that require not only ubiquitous communication but also user localization capabilities. Consequently, research interest in advanced localization methods has surged, among which holographic multiple-input multiple-output (HMIMO)-based localization methods are expected to play a critical role due to its potential to achieve high localization accuracy. Specifically, HMIMO is an electromagnetic surface containing a massive number of antennas or reconfigurable metamaterial elements\cite{b48}\cite{b56}. Due to the large number of tunable elements in the surface, HMIMO can precisely manipulate wireless beams to accurately locate users.

\subsection{Literature Review and Motivation}

Traditional HMIMO-based localization works mainly fall into two categories: far-field~(FF)\cite{b82,b77, b85,b14, b57, b81, b59} and near-field~(NF) localization schemes~\cite{b6,b86,b61,b58,b80,b83}. In the FF region, the distances between the HMIMO and users are sufficiently large, and the signal channel is effectively described using a plane wave model. In \cite{b82}, the fundamental bounds and localization feasibility condition are examined for FF localization in orthogonal frequency division multiplexing (OFDM) systems aided by HMIMO. The authors in~\cite{b77} considered a single-input-single-output~(SISO) HMIMO-aided localization system with imperfect HMIMO geometry information, and derived the misspecified Cramér-Rao bound (CRB). In \cite{b85}, the positioning estimation error and orientation estimation error are investigated for a HMIMO-aided multiple-input multiple output system. In \cite{b14}, the authors analyzed the CRB for a localization scenario that both the line-of-sight~(LOS) path and the reflected path via the HMIMO exist simultaneously. Two HMIMO phase design methods were proposed and a maximum likelihood (ML) direction of arrival (DOA) estimator was designed in \cite{b14}. In~\cite{b57}, the authors considered FF user localization aided by millimeter wave (mmWave) HMIMO, and a ML location estimator and two sub-optimal estimators with lower complexity were derived. In \cite{b81}, the localization method leveraged the dynamic regulation of HMIMO to cancel non-line-of-sight signals at the receiver and preserve LOS signals for precise indoor localization. HMIMO-based localization without base stations~(BSs) was also investigated in \cite{b59}, where the user transmitted the signal and received the signal reflected from the HMIMO.
% In \cite{b65}, the authors considered the scenario that both the line-of-sight~(LOS) path and the reflected path via the HMIMO exist simultaneously and proposed a low-complexity ML estimator. 

HMIMO-based NF localization schemes have attracted much attention recently due to the increasing interest in large-scale HMIMO. Specifically, to improve beam control capability, the scale of HMIMO is enlarged, leading to the expansion of the NF region\cite{b5}. Different from the FF case, the spherical wave model has to be employed to characterize the NF channel because the signal transmission characteristics of the NF are determined by both the range and angle of the user location. 
% bound
The authors in \cite{b6} and \cite{b86} considered a SISO localization scenario for NF user, and the CRB is derived.
% method
In \cite{b61}, a HMIMO functioning as a lens receiver was utilized to achieve NF localization. The authors in~\cite{b58} developed a received signal strength-based localization method for NF users and a RIS phase shift adjustment method to enhance accuracy. In~\cite{b80}, the authors utilized light detection and ranging to estimate scatter locations and improve NF user localization accuracy. A signaling and positioning method is proposed in~\cite{b83}, and the coverage capability under different levels of obstruction of the HMIMO is analyzed.

% The ability of beamforming increases with the number of elements in the HMIMO. Hence a larger HMIMO can enhance localization accuracy. However, the large aperture of the HMIMO also leads to the expansion of the near-field (NF) region\cite{b5}. The NF is the region where the transmission characteristic of the signal is related to the range and angle of the user location. Therefore, it is necessary to model the NF channel using the spherical wave model. On the other hand, the far-field (FF) is where the signal channel is only related to the angle of the user and can be modeled by the plane wave model. Due to the different signal transmission characteristics of the NF and FF, these two regions need to be treated differently.

% Existing methods primarily focus on the pure NF or FF region. In \cite{b50}, the author derived the error bounds for a reconfigurable intelligent surface~(RIS)-aided 3D NF localization and orientation estimation performance of synchronous and asynchronous signaling schemes. In \cite{b49}, a RIS-aided millimeter wave (mmWave) multiple-input-multiple-output (MIMO) radar system for multi-target localization was proposed, and the simulation of a blocked line of sight (LOS) scenario was provided.

In practical scenarios, users may exist in both the NF and FF regions of the HMIMO. Existing algorithms for pure NF or FF users are therefore not always applicable. Specifically, FF localization methods suffer from low localization accuracy for NF users due to the significant error in describing NF channels using the plane wave model. Meanwhile, NF localization techniques are not ideal for FF users because the coupling of range and angle in the spherical model brings extra complexity, which has two main effects: first, it prolongs the running time of the algorithm; second, the more complex NF model hinders convergence of the method and leads to accuracy degradation. In addition, the localization gain for the FF users brought by the NF model is limited. Hence, a localization scheme is required to adapt to both the NF and FF users.

\subsection{Main Contributions}
In this paper, a HMIMO-based localization scheme is proposed for users in both NF and FF regions. Specifically, in the considered scenario, a base station~(BS) and a reconfigurable intelligent surface~(RIS) cooperate to locate multiple users in a multipath environment. The RIS is widely acknowledged as a low-cost implementation of HMIMO that can create a customized reflection beam by varying the phase shifts of its reflection elements. In the localization process, the RIS reflects the signals emitted from several single-antenna users to the BS. Then, the BS uses the received signal to estimate the locations of the users in both NF and FF regions. The phase shifts of the RIS are optimized based on the estimated locations of the users. The location estimation and RIS phase shift optimization are performed iteratively to improve localization accuracy.

% and then based on the optimized phase shifts, the user locations are estimated again using received signals. In this way, the localization accuracy can be iteratively improved.

Several challenges in this scenario need to be addressed. \textit{First}, it is unknown whether each user or scatter is in the NF or the FF region, leading to difficulty in designing the localization algorithm. \textit{Second}, the channels of different users are coupled due to shared RIS phase shifts and common scatters. To tackle the above challenges efficiently, a hybrid NF and FF localization algorithm is proposed, in which we first sample the NF and FF regions separately, and then compare the received signals with the reconstructed signals at the sampled locations to locate the users and scatters. The locations of the users and scatters are jointly estimated by considering the channel coupling effect. Then, a RIS phase shift optimization problem is formulated to minimize the sum of the Cramér-Rao bounds (CRBs) of the user localization errors. A complex circle manifold-based method is proposed to solve the optimization problem with the coupled CRBs. 

Our contributions can be summarized as follows.

\begin{itemize}
	\item We consider a hybrid NF and FF localization scenario, where the BS locates multiple users in hybrid regions with the aid of an RIS. A hybrid NF and FF localization framework is proposed, where the localization of the users and the RIS phase shifts optimization are iteratively conducted to improve localization accuracy.

	\item Based on the principle of minimizing localization loss, the location estimation problem is formulated and a multi-user localization algorithm is developed, which localizes users in the hybrid NF and FF region under the channel coupling effect.
	
	\item We formulate the RIS phase shift optimization problem to minimize the sum of CRBs of the location estimation errors. A complex circle manifold-based algorithm is then devised to obtain optimal RIS phase shifts.

	\item We compare the performance of the proposed approach and other RIS-enabled localization techniques. Simulation results show that the proposed approach outperforms other methods by more than 30\% in the root mean square error~(RMSE) of angle estimation for both NF and FF users and by 60\% in the RMSE of range estimation for NF users given $-10\leq\text{SNR}\leq0$.
\end{itemize}

\subsection{Organization}
The rest of this paper is organized as follows. The localization scenario, the signal model, and the localization protocol are provided in Sec.~\ref{section:sm}. In Sec.~\ref{section:location}, we propose the localization algorithm. Then, the RIS phase shift optimization algorithm is designed in Sec.~\ref{section:prob}. The performance of the proposed method is analyzed in Sec.~\ref{section:peran}. In Sec.~\ref{section:simu}, we present the simulation results and discussions. Finally, conclusions are drawn in Sec.~\ref{sec:con}.

\textit{Notation:} The transpose and conjugate transpose are denoted by $(\cdot)^T$ and $(\cdot)^H$, respectively. We use $\Vert \bm{x} \Vert_2$ and $\Vert \bm{x} \Vert_0$ mean the $l_2$ and $l_0$ norm of vector $\bm{x}$, and $\vert \cdot \vert$ denotes the number of elements in a set. Unbolded $x$, bold lower-case $\bm{x}$ and bold upper-case $\bm{X}$ denote scalar, vector and matrix, respectively. The collection of all complex numbers is denoted by $\mathbb{C}$. The Hadamard product is represented by $\odot$.

\section{System Model}
\label{section:sm}
In this section, we first introduce the scenario of multi-user localization. Then, we describe the signal model and introduce a localization protocol for the proposed scenario.

\begin{table}
	\centering
	\caption{Major Notations}
	\normalsize
	\resizebox{85mm}{!}{
		\begin{tabular}{l l} 
		\toprule Notation & Description\\ 
		\midrule 
		$N$ & Number of RIS elements  \\
            $K$ & Number of users \\
            $L$ & Number of scatters \\
            $C$ & Number of localization cycles \\
            $\bm{\beta}$ & RIS phase shifts \\
            $\bm{B}^{(c)}$ & RIS phase shifts for $c$ cycles \\
            $\bm{h}_k$  & Channel between the $k$-th user and the BS\\
            $\bm{h}^A_k$ & Channel between the RIS and the BS\\
            $\bm{h}^t_k$ & Channel between the RIS and the $k$-th user \\
            % $\bm{h}^t_{kl}$ & $l$-th path in the user-RIS channel \\
            $s_k$ & The transmitted signal by the $k$-th user \\
            $\bm{p}_k$ & Location of the $k$-th user \\
            $\bm{Q}$ & Locations of all scatters  \\
            $\bm{g}^{(c)}_k$ & The received signal of the $c$-th cycle \\
            $\bm{Z}$ & The sampled locations \\
            $\bm{F}$ & Atom channel in the NF and FF regions \\
            $\bm{u}_k$ & Gains of the atom channels for the $k$-th \\
            $\bm{u}_k^{\bm{P}}$ & The direct path component in $\bm{u}_k$ \\
            $\bm{u}_k^{\bm{Q}}$ & The scattering path component in $\bm{u}_k$ \\
            $\bm{\Lambda}^k$ & Atom signals for the $k$-th user \\
            $\bm{D}_{NF}$, $\bm{D}_{FF}$ & NF and FF region \\
            $D_R$, $D_\theta$, $D_\phi$ & CRBs for range and angle \\
		\bottomrule 
		\end{tabular}
	}
\end{table}

\subsection{Localization Scenario}

\begin{figure}
	\centering
	\includegraphics[scale=0.25]{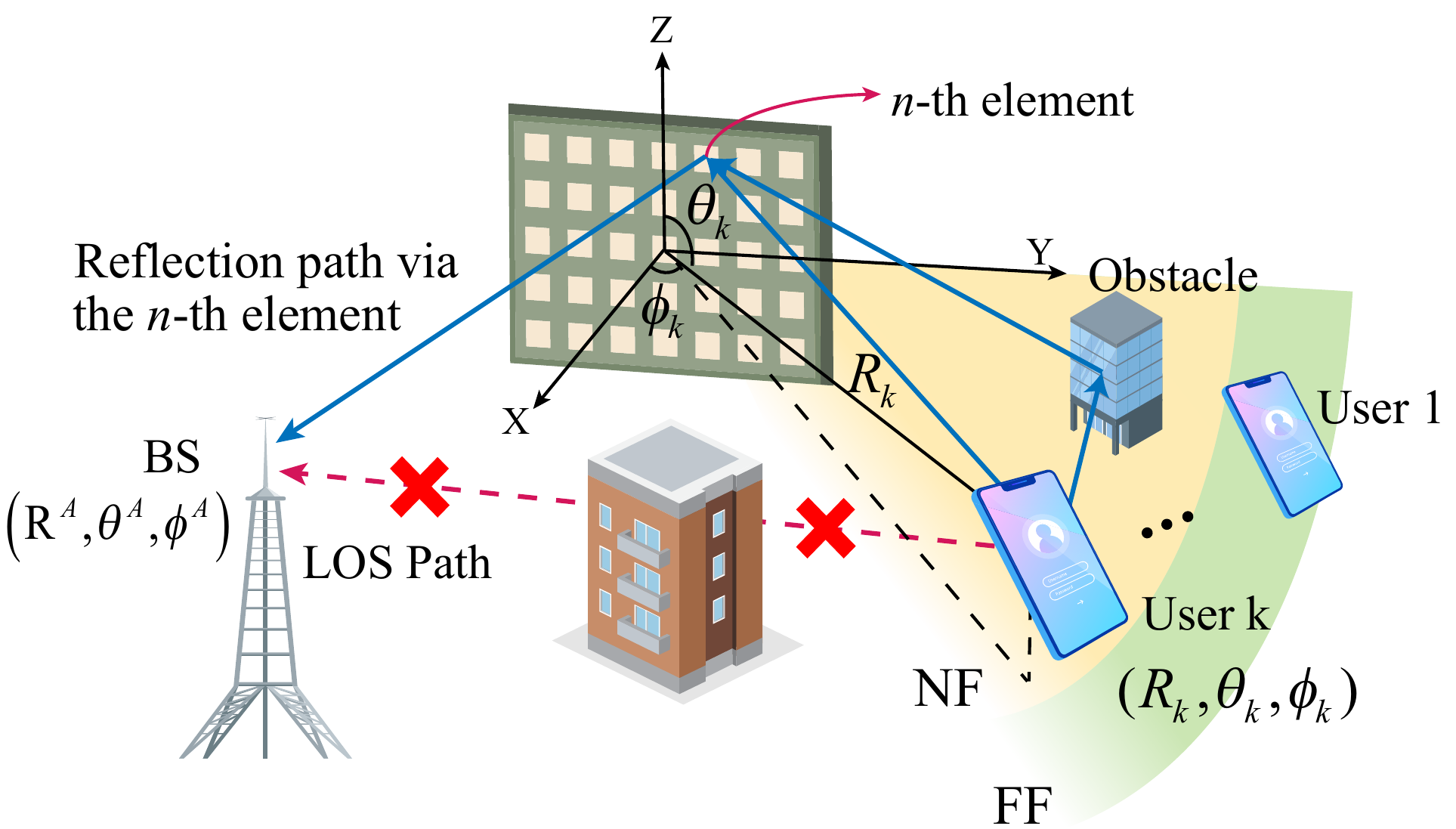}
	\caption{RIS-enabled multi-user localization system.}
\end{figure}

As shown in Fig. 1, we consider a localization scenario consisting of $K$ users, a RIS, and a BS\footnote{In this work, we use reflective reconfigurable intelligent surface~(RIS) as an example to implement HMIMO, and it can be easily replaced with other surfaces, such as transmissive RIS and reconfigurable holographic surface.}. The users are stationary or low-speed compared with the localization time. Users and the BS are each equipped with a single antenna. The RIS containing $N=N_1\times N_2$ elements is placed on the Y-Z plane, where the center of the RIS is at the origin of the coordinate system. There also exist obstacles, which reflect the signals and create multipath channels. We assume the line-of-sight path between each user and the BS is blocked, and there only exist the reflection paths via the RIS. 

During the localization process, each user sends a narrowband signal to the RIS, and the RIS reflects it to the BS. To avoid multi-user interference, frequency division multiplexing~(FDM) is for different users, i.e. the $k$-th user sends the signal over frequency $f_k$, where $f_k \ne f_k'$ ($k \ne k'$). The BS then estimates the locations of the users using the received signals. We assume the locations of the RIS and the BS are known, while the locations of the users and scattering centers in the obstacles~\cite{b33} are unknown. Whether the users are located in the NF or the FF regions of the RIS is also unknown.

\subsection{Signal Model}

In the uplink transmission of the $k$-th user, the signal $y_k$ received by the BS is given by
\begin{equation}
	y_k=\bm{\beta}^T\bm{h}_k(\bm{p}_k,\bm{Q})s_k+\epsilon,
\end{equation}
where $\bm{\beta}=[\beta_{1},\beta_{2},...,\beta_{N}]^T \in \mathbb{C}^{N\times 1}$ is the phase shift vector of the RIS which is shared for all $K$ users, $\beta_{n}$ represents the phase shift of the $n$-th RIS element with $|\beta_{n}|=1$, $s_k \in \mathbb{C}^{1\times 1}$ represents the narrowband signal transmitted by the $k$-th user, and $\epsilon$ represents received Gaussian noise with noise power $\sigma^2$ and mean $0$. The location of the $k$-th user is given by $\bm{p}_k=[R_k,\theta_k, \phi_k]^T$, with $R_k$, $\theta_k$, $\phi_k$ being the range, polar angle, and azimuth angle, respectively. Here $\bm{Q} = [\bm{q}_1,...,\bm{q}_{L_{max}}] \in \mathbb{C}^{3\times L_{max}}$ denotes the locations of the scatters, where the number of scatters is assumed to be less than $L_{max}$, and the location of the $l$-th scatter is given by $\bm{q}_l = (r_l, \vartheta_l, \varphi_l)^T$. The cascaded channel between the $k$-th user and the BS $\bm{h}_k(\bm{p}_k,\bm{Q})$ is given by to $\bm{h}^A_k \odot \bm{h}_k^t(\bm{p}_k,\bm{Q})$, where $\bm{h}^A_k$, $\odot$, and $\bm{h}_k^t(\bm{p}_k,\bm{Q})$ denote the RIS-BS channel for the $k$-th user, the Hadamard product, and the channel between the RIS and the $k$-th user, respectively. The channel between the $k$-th user and the RIS is given by~\cite{b36}
\begin{align}
	\label{channelmodel}
	\bm{h}_k^t(\bm{p}_k,\bm{Q}) = \bm{h}^t_{k0}(\bm{p}_k) + \sum_{l=1}^{L} \bm{h}^t_{kl}(\bm{p}_k,\bm{q}_l),
\end{align}
where $\bm{h}^t_{kl}(\bm{p}_k,\bm{q}_l)$ represents the $l$-th path between the RIS and the $k$-th user, and $L$ is the number of scatters in the scenario. We assume $l = 0$ is associated with the direct path between the $k$-th user and RIS, while $l \neq 0$ are the paths via scatters. Since users or scatters could be in either NF or FF regions of the RIS, the channels in these two cases are modeled separately.

\subsubsection{Channel Models for the NF Region}
\label{subsubsection:nf}
When the user or the scatter is located in the NF region, the signal received by the RIS is described using the spherical wave model and cannot be approximated as a plane wave for the RIS. Thus, the direct user-RIS channel is modeled by\cite{b34}
\begin{align}
	\label{nfdirect}
	\bm{h}^t_{k0}(\bm{p}_k) = \alpha_k \bm{b}(R_k, \theta_k, \phi_k), \ \bm{p}_k \in \bm{D}_{NF},
\end{align}
where $\alpha_k$ is the channel gain including the effects of path loss and the directivity of the RIS elements\cite{b20}, and $\bm{b}$ is the steering vector in the NF region, given as
\begin{flalign}
	\! \!\! \bm{b}(R_k, \theta_k, \phi_k) \! =  \! \! \left[  \exp \! \left( \! -j\dfrac{2\pi}{\lambda}d_{k1}^t \! \right)\!,  \dots ,  \exp \! \left( \! -j\dfrac{2\pi}{\lambda}d_{kN}^t  \! \right) \! \right]^T \!\!\!\!. \!\!
\end{flalign}
Here, $d_{kn}^t$ is the distance between the $k$-th user and the $n$-th RIS element, and $\bm{D}_{NF}$ represents the NF region. Based on \cite{b5}, the NF region is defined as the region where the maximum phase error between the phase shift calculated under plane wave approximation and the real phase shift is more than $\pi/8$. Mathematically, the NF region is given by $ \bm{D}_{NF}=\{\bm{p} \vert \Delta \varphi(\bm{p}) > {\pi}/{8} \}$, where $\Delta \varphi(\bm{p})$ is the maximum phase error across all the RIS elements, given by
\begin{flalign}
    \!\Delta \varphi(\bm{p}) \! = \! \mathop{\max}\limits_{n}\frac{2\pi}{\lambda} \! \left(d_{n}^t - (R-y_n \sin\theta \sin\phi - \! z_n \cos\theta)\right),
\end{flalign}
where $d_{n}^t$ is the distance to the $n$-th RIS element, and $(0, y_n, z_n)$ is the coordinate of the $n$-th RIS element.

Similarly, we model the paths via the scatters. Specifically, when the $l$-th scatter is located in the NF region, the path between the $l$-th scatter and the RIS is given by
\begin{align}
	\label{nfsc}
	\bm{h}_{kl}^t(\bm{p}_k,\bm{q}_l) =  \alpha_{kl} \bm{b}(r_l, \vartheta_l, \varphi_l), \ \bm{q}_l \in \bm{D}_{NF},
\end{align}
where $\bm{b}(r_l, \vartheta_l, \varphi_l) = [\exp(-j\frac{2\pi}{\lambda}d_{l1}^t),...,\exp(-j\frac{2\pi}{\lambda}d_{lN}^t)]^T$, and $d_{ln}^t$ is the distance between $l$-th scatter and $n$-th element of the RIS. Here $\alpha_{kl}$ is the channel gain of the $l$-th path.

\subsubsection{Channel Models for the FF Region}
\label{subsubsection:ff}
When the user is located in the FF region, we adopt the plane wave approximation to model the received signals. The direct user-RIS channel is modeled by\cite{b34}
\begin{align}
	\label{ffdirect}
	\bm{h}^t_{k0}(\bm{p}_k) = \alpha_k \bm{a}(\theta_k, \phi_k),  \ \bm{p}_k \in \bm{D}_{FF},
\end{align}
where $\bm{a}(\theta_k, \phi_k)$ is the FF steering vector and is given as~\cite{b29}~\cite{b89}
\begin{align}
	\bm{a} & (\theta_k, \phi_k) = \left[\exp \left(-j\dfrac{2\pi}{\lambda}(-y_1\sin\theta_k \sin\phi_k -z_1 \cos\theta_k)\right),\right.\nonumber\\
	& \left....,\exp \left(-j\dfrac{2\pi}{\lambda}(-y_N\sin\theta_k \sin\phi_k -z_N \cos\theta_k)\right) \right]^T.
\end{align}
In (\ref{ffdirect}), $\bm{D}_{FF}$ represents the FF region, which is defined as the area where the maximum phase error is less than $\pi/8$, i.e., $\bm{D}_{FF}=\{\bm{p}\vert \Delta \varphi(\bm{p}) < \pi/8 \}$. When the $l$-th scatter is located in the FF region, the path between the $l$-th scatter and the RIS can be modeled by
\begin {align}
	\label{ffsc}
	\bm{h}_{kl}^t(\bm{p}_k,\bm{q}_l) =  \alpha_{kl} \bm{a}(\vartheta_l, \varphi_l), \ \bm{q}_l \in \bm{D}_{FF}.
\end{align}

\subsection{Localization Protocol}
\label{subsection:LP}
\begin{figure}
	\centering
	\includegraphics[scale=0.35]{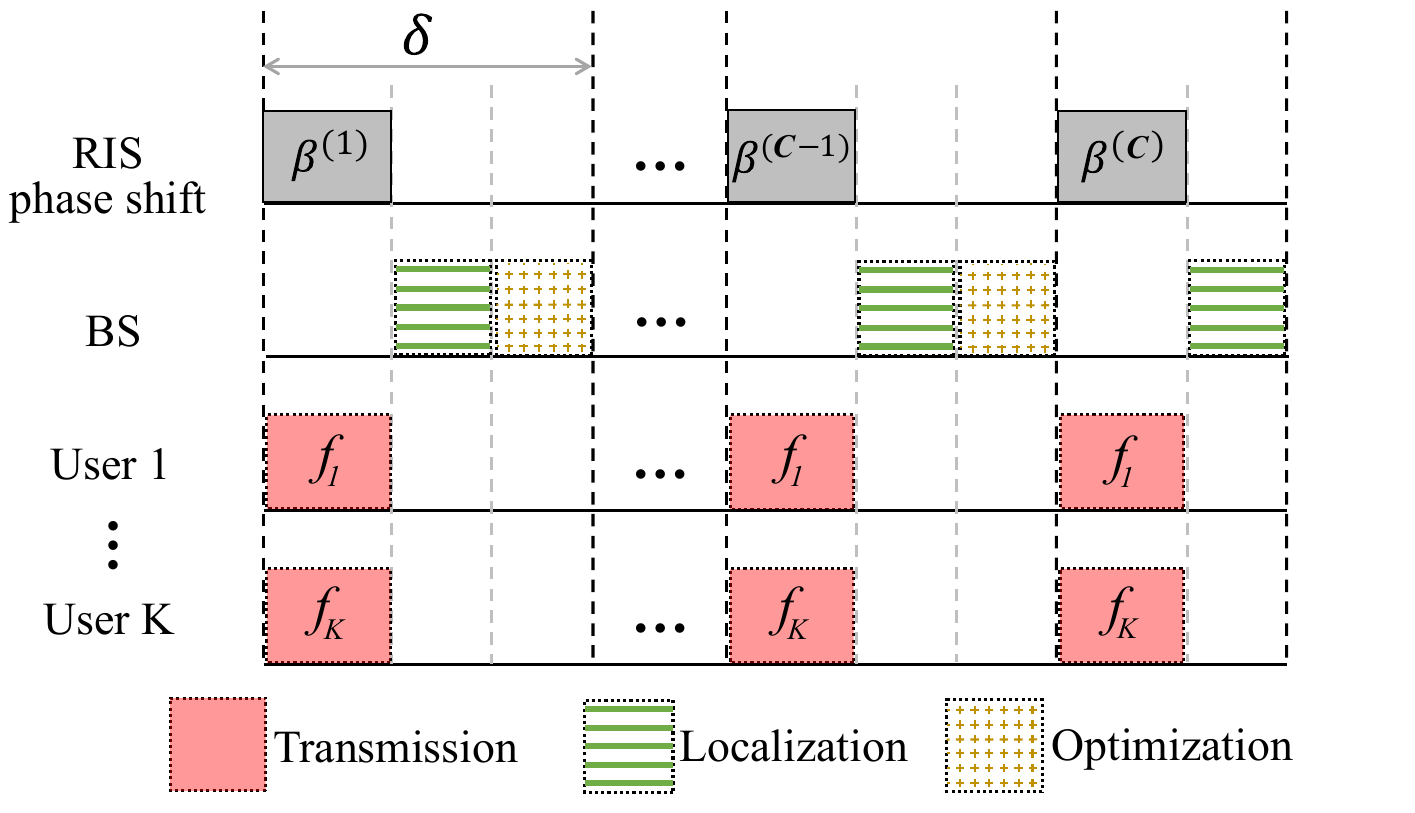}
	\caption{RIS-enabled hybrid NF and FF source localization protocol.}
	\label{fig:2}
\end{figure}

We propose an RIS-enabled hybrid NF and FF localization protocol, which improves localization accuracy by iteratively localizing the users and optimizing the RIS phase shifts based on the estimated locations. The localization process is divided into $C$ cycles with cycle duration being $\delta$. Note that cycle duration $\delta$ can be adjusted for different systems. Each cycle contains three steps: transmission, localization, and optimization. The process of the localization protocol is illustrated in Fig. \ref{fig:2}.

\subsubsection{Transmission}
In this step, the users send signals to the RIS, which are reflected to the BS. Let $\bm{g}^{(c)} = [g^{(c)}_1,...,g^{(c)}_K]^T \in \mathbb{C}^{K\times 1}$ denote the signal received by the BS in the $c$-th cycle. The RIS phase shifts are set randomly in the first cycle, while in the following $C-1$ cycles, the phase shifts are selected based on the optimization results in the previous cycle, which will be described in the optimization step.

\subsubsection{Localization}
\label{subsubsection:localization}
In the next step, the BS estimates users' locations using the received signals. Specifically, in the $c$-th cycle, based on the received signals $\bm{G}^{(c)}=[\bm{g}^{(1)},...,\bm{g}^{(c)}]^T  \in \mathbb{C}^{K\times c}$, we determine whether the users and the scatters are located in the NF or the FF region and jointly estimate their locations $\bm{P}^{(c)} = [\bm{p}_1^{(c)},...,\bm{p}_K^{(c)}]$ and $\bm{Q}^{(c)} = [\bm{q}_1^{(c)},...,\bm{q}_L^{(c)}]$. This step focuses on the problem of accurately locating users given the received signals and the RIS phase shifts. The details of the localization algorithm are introduced in Sec.~\ref{section:location}.
% We also jointly estimate the locations of the scatters $\bm{q}^{(c)} = [\bm{q}_1^{(c)},...,\bm{q}_L^{(c)}]$ to enhance localization accuracy of the users. 

% \subsubsection{Response}
% In next $\delta_R$ seconds, the users are required to report their estimated locations to the BS. Then, these locations will be utilized for RIS phase shifts optimization.

\subsubsection{Optimization}

In the rest time of the $c$-th cycle, the optimal RIS phase shifts $\bm{\beta}^{(c+1)}$ are selected according to the estimated locations in the former step. Note that this step is not executed in the last cycle. This step focuses on the task of calculating the optimal RIS phase shifts for the next cycle based on the user localization results of the previous step. The details of optimization are introduced in Sec.~\ref{section:prob}.
% $\bm{B}^{(c+1)} = [\bm{\beta}^{(c+1)}_1,...,\bm{\beta}^{(c+1)}_K]$

% In summary, two parts of the high-level problem need to be solved: localization and optimization, which iterate with each other to achieve optimal RIS phase shifts and higher localization accuracy. Two algorithms are designed to solve these problems respectively.

\section{RIS-enabled Hybrid Near-field and Far-field Localization}
\label{section:location}

In this section, we first formulate the localization problem and then propose a localization algorithm to solve the formulated problem.

\subsection{Localization Problem Formulation}
\label{local:formu}
% Note that in the multi-user scenario, the scattering paths of different users are coupled with each other. For example, signals transmitted by different users may pass through the same scatter. Therefore, conducting multi-user joint localization can identify the same scatter, and thereby improving localization accuracy compared to single-user localization. Hence, the localization problem is modeled as a joint localization problem.
% Note that while the received signal is influenced by both the user and scatter locations, our focus is solely on the user locations. Consequently, we only optimize the user locations in order to achieve the minimum localization loss. 

% Different from standard sparse recovery approach, we focus on the locations of users and want to only minimize the estimation error of user locations. Thus, the localization problem is formulated as follows to separate the parameters of interest and those of non-interest.

% The user-RIS channel is inherently determined by the spatial coordinates of users and scatters. Therefore, we are able to recover user locations from channel estimates. 
We formulate the multi-user localization problem to minimize the user localization loss, which is defined as the sum of the $l_2$-norm of the residual between the received signals and the signals reconstructed given user locations. Thus, the localization problem is formulated as
\begin{subequations}
	\begin{align}
		% \label{prob:1}
			\text{P1}: &\mathop{\min}_{\bm{P}}  \sum_{k=1}^{K} \left\Vert \bm{g}^{(c)}_k - (\bm{B}^{(c)})^T\bm{h}_k(\bm{p}_k,\overline{\bm{Q}}) s_k \right\Vert_2^2 , \label{prob:1a}\\
			% \text{P1}: &\mathop{\min}_{\bm{P},\bm{Q}} \Vert \bm{G}^{(c)}-(\bm{B}^{(c)})^T \bm{H}(\bm{P},\bm{Q}) s \Vert_F^2 \\
			&s.t. \ \ \bm{p}_k \in \bm{D}_{NF} \cup \bm{D}_{FF},\forall k,
	\end{align}
\end{subequations}
where $\bm{g}^{(c)}_k = [g_k^{(1)},...,g_k^{(c)}]^T \in \mathbb{C}^{c\times 1}$ is the received signals of the $k$-th user in the previous $c$ cycles, i.e., the $k$-th row of $\bm{G}^{(c)}$. Here $\bm{B}^{(c)} = [\bm{\beta}^{(1)},...,\bm{\beta}^{(c)}]  \in \mathbb{C}^{N\times c}$ is the phase shifts of the RIS in the previous $c$ cycles, $\bm{P} = [\bm{p}_1,...,\bm{p}_K]$ is the locations of the users, $\overline{\bm{Q}}$ is the ground truth of scatter locations. It can be observed from (\ref{prob:1a}) that the objective function is affected by the accuracy of user location estimation. By solving (P1), we can minimize the estimation error of the user location and eliminate the interference caused by the scattering paths.

% $\Vert \cdot \Vert_F$ denotes the Frobenius norm.
% where $\bm{H}(\bm{P},\bm{Q})$ is the cascaded channel for $K$ users and given by $\bm{H}(\bm{P},\bm{Q}) = [\bm{h}_1(\bm{p}_1,\bm{Q}),...,\bm{h}_K(\bm{p}_K,\bm{Q})]$.

Compared to single-user localization, multi-user localization is more challenging because the channels of different users are coupled with each other. Specifically, the signals transmitted by different users at different frequencies may pass through the same scatter. Thus, the localization loss of each user, i.e., a term in the summation (\ref{prob:1a}), is affected by not only the location of each user but also the scatter locations $\bm{Q}$. This is different from the single-user localization case where the locations of a user and the scatters are optimized to minimize only one localization loss term and the coupling of the localization losses among different users
is not considered.

Due to the non-convex nature of the above problem (P1), conventional algorithms like gradient descent can easily fall into local minima. To solve it effectively, a location estimation method by modifying the grid search method is proposed to perform a global search. Specifically, we first sample the search domain in NF and FF regions to create a grid map of the sampled candidate locations, denoted by $\bm{Z} = [\bm{Z}_{near}, \bm{Z}_{far}]$, where $\bm{Z}_{near}$ and $\bm{Z}_{far}$ are the sampled candidate locations in the NF and FF, respectively. Then, the most suitable sample locations are selected to minimize the localization loss. 

Considering the different signal models in the NF and FF regions, the locations in these two regions are sampled in different ways, which are elaborated in the following.

\textbf{NF case:} In the NF region, $R, \theta, \phi$ are uniformly sampled with sampling spacings $\Delta R,\Delta \theta=\pi/N_{\theta}, \Delta \phi=\pi/N_{\phi}$, respectively, where $N_{\theta}$ and $N_{\phi}$ are predetermined parameters. Thus, $S$ sampled candidate locations in the NF are obtained.
\begin{align}
	\bm{Z}_{near} = \left[\left[R_1,\theta_1,\phi_1\right]^T, ..., \left[R_S,\theta_S,\phi_S\right]^T\right].
\end{align}
We define the atom channels as the user-RIS channels or the scatter-RIS channels between the RIS and every possible candidate location of the users or the scatters. The atom channel in the NF region is given by\cite{b13}
\begin{equation}
	\label{atomNF}
	\bm{F}_{near} = \left[\bm{b}(R_1,\theta_1,\phi_1),...,\bm{b}(R_S,\theta_S,\phi_S)\right] \in \mathbb{C}^{N\times S} ,
\end{equation}
where $\bm{b}(R_i,\theta_i,\phi_i)$ is the NF steering vector given the location $[R_i,\theta_i,\phi_i]^T$.

%  Unlike the NF case, the range of the FF user cannot be estimated because the angles of arrival for different RIS elements are very close.
\textbf{FF case:} In the FF region, only the angles $\theta, \phi$ are sampled. The range $R$ is not sampled because it does not affect the FF steering vector and it cannot be estimated similarly to NF case. The primary reason lies in the angular relationships between the user and the RIS elements. Specifically, in the NF, the angles between the user and each RIS element are different. This leads to different phase shift changes for paths from the user to each RIS element and varied received signals when the range varies, which allows for the estimation of the range $R$. In contrast, for FF users, since the angles between the user and each RIS element are almost the same, when the range changes, the phase shift changes for the path from the user to each RIS element are the same. This uniform phase shift leads to an overall phase change in the received signal, hence range estimation cannot be achieved.

As a result, the number of candidate locations and the algorithm complexity are significantly reduced compared with NF. We apply the same angle sampling methods in the FF as in the NF. The atom channel of the FF region is given by~\cite{b12}
\begin{align}
	\label{atomFF}
	\bm{F}_{far} = [\bm{a}(\theta_1,\phi_1),...,\bm{a}(\theta_{N_{\theta}N_{\phi}},\phi_{N_{\theta}N_{\phi}})] \in \mathbb{C}^{N\times N_{\theta}N_{\phi}},
\end{align}
where $\bm{a}(\theta_i,\phi_i)$ is the FF steering vector of the angle $[\theta_i,\phi_i]$.

\begin{figure}
	\centering
	\includegraphics[scale=0.28]{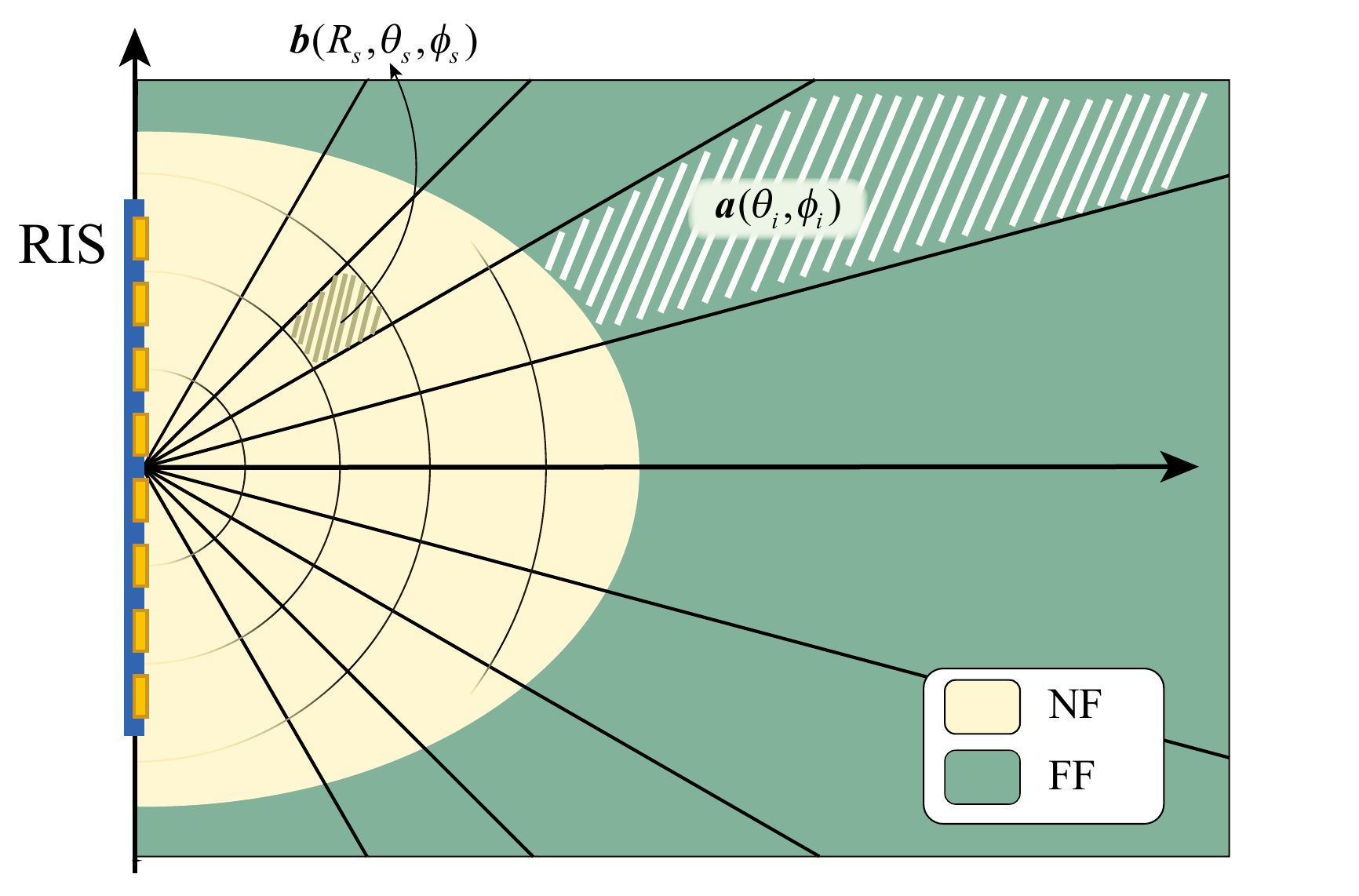}
	\caption{Illustration of the sampled locations at $\theta = \pi/2$.}
	\label{fig:3}
\end{figure}

We define $\bm{F}=[\bm{F}_{near},\bm{F}_{far}] \in \mathbb{C}^{N\times M}$ as the atom channels for the hybrid field, where $M = S + N_{\theta}N_{\phi}$. For better illustration, the sampled locations at $\theta = \pi/2$ plane for both the NF and FF regions are shown in Fig. \ref{fig:3}.

% To solve it effectively, we propose a location estimation algorithm by modifying the grid search method in order to perform global search. Specifically, we sample the search domain and create a grid map of the sampled candidate locations, denoted by $\bm{Z}$. Note that for the NF region, $R, \theta, \phi$ are all sampled because the range and angles are coupled in the NF region. While for the FF region, the range $R$ is not sampled because the range only affects the overall channel amplitude, which can be addressed by linear regression. Besides, unlike the NF case, the range cannot be estimated for the FF user because the angles of arrival at all RIS elements are approximately the same. In this way, we can decrease the number of candidate locations and then decrease the algorithm complexity. The specific sampling method of candidate locations $\bm{Z}$ will be explained in detail in the following.

% We define $\bm{F}=[\bm{F}_{near},\bm{F}_{far}]$ as the atom channels, which are the user-RIS channels between the RIS and every possible candidate locations of the user. $\bm{F}_{near}$ and $\bm{F}_{far}$ are the atom channels in the NF and FF regions, which will be explained in detail in the following. 
Based on the channel model (\ref{channelmodel}) and the atom channels (\ref{atomNF}) and (\ref{atomFF}), each user-RIS channel can be approximated as a linear combination of multiple atom channels. We define $\bm{u}_k \in \mathbb{C}^{M \times 1}$ as the gains of the atom channels for the $k$-th user-RIS channels, which can be decomposed into two components: $\bm{u}^{\bm{P}}_k$ for the direct path, and $\bm{u}^{\bm{Q}}_k$ for the scattering path. Both $\bm{u}^{\bm{P}}_k$ and $\bm{u}^{\bm{Q}}_k$ are unknown, and our aim is to accurately estimate $\bm{u}^{\bm{P}}_k$, which contains user locations information. Specifically, $u_{ik}^{\bm{P}}$, the $i$-th element of $\bm{u}^{\bm{P}}_k$, denotes the gain of the $i$-th atom channel in the direct path between the $k$-th user and the RIS. If $u_{ik}^{\bm{P}} = 0$, it indicates that the $k$-th user is not at location $\bm{z}_i$, the ${i}$-th column of the candidate locations $\bm{Z}$. Otherwise, the $k$-th user is located at $\bm{z}_i$, and the estimated gain is $u_{ik}^{\bm{P}}$. In contrast, we do not care about the estimation accuracy of $\bm{u}^{\bm{Q}}_k$, because it is only influenced by scatter locations.

% We define $\bm{U}(\bm{P},\bm{Q})$ as the gains of all the atom channels to constitute $K$ user-RIS channels. $u_{ij}$ is the $(i, j)$-th element of $\bm{U}(\bm{P},\bm{Q})$, denoting the gain of the $i$-th atom channel in the user-RIS channel of the $j$-th user, i.e. $\alpha_k$ or $\alpha_{kl}$ in (\ref{nfdirect}), (\ref{nfsc}), (\ref{ffdirect}) and (\ref{ffsc}). If $u_{ij} = 0$, it indicates that there is no user or scatter at location $\bm{z}_i$, which is the ${i}$-th column of the candidate locations $\bm{Z}$. Otherwise, there is a user or scatter at location $\bm{z}_i$, and the corresponding gain is $u_{ij}$. 

The $k$-th user-RIS channels are thus given as
% Each nonzero element in $\bm{U}(\bm{P},\bm{Q})$ corresponds to the location of the users or the scatters.
\begin{align}
	% \bm{H}^t(\bm{P},\bm{Q}) &= [\bm{h}_1^t(\bm{p}_1,\bm{Q}),...,\bm{h}_K^t(\bm{p}_K,\bm{Q})] \nonumber\\
	% &= \bm{F}\bm{U}(\bm{P},\bm{Q}).
	\bm{h}^t_k(\bm{p}_k,{\bm{Q}}) &= \bm{F} \bm{u}_k = \bm{F}(\bm{u}^{\bm{P}}_k + {\bm{u}^{\bm{Q}}_k}).
\end{align}
% where $\bm{u}^{\bm{P}}_k$ and $\bm{u}^{\bm{Q}}_k$ are the $k$-th column of $\bm{U}^{\bm{P}}$ and $\bm{U}^{\bm{Q}}$, respectively.
The cascaded channel of the $k$-th user is given by 
\begin{align}
	\bm{h}_k(\bm{p}_k,{\bm{Q}}) = \bm{h}^A_k \odot \bm{h}_k^t = \text{diag}\{\bm{h}^A_k\}\bm{F}(\bm{u}^{\bm{P}}_k + {\bm{u}^{\bm{Q}}_k}). 
\end{align}

Since the number of users and scatters are generally much smaller than the number of atom channels, $\bm{u}_k$ can be directly solved base on traditional sparse recovery method like orthogonal matching pursuit~(OMP). However, the user location estimated in this way is not accurate enough due to interference of scattering paths. To mitigate this interference and improve user localization accuracy, (P1) is approximated as
{
	\begin{subequations}
		\label{eq:6}
		\begin{align}
			\text{P1'}: \mathop{\min}_{\bm{u}^{\bm{P}}_k,\forall k}  &\sum_{k=1}^{K} \left\Vert \bm{g}^{(c)}_k - \bm{\Lambda}^k(\bm{u}^{\bm{P}}_k + \overline{\bm{u}^{\bm{Q}}_k})  \right\Vert_2^2 , \\
			s.t. \ & \left\Vert \bm{u}^{\bm{P}}_k \right\Vert_0 = 1, \forall k, 
		\end{align}	
	\end{subequations}
}%
where $\overline{\bm{u}^{\bm{Q}}_k}$ is the ground truth of ${\bm{u}^{\bm{Q}}_k}$. Here, $\bm{\Lambda}^k = (\bm{B}^{(c)})^T \text{diag}\{\bm{h}^A_k\}\bm{F} s_k \in \mathbb{C}^{c\times M} $ is the atom signals, which are the linear transformation of the atom channels $\bm{F}$ by considering the effect of RIS phase shift $\bm{B}^{(c)}$, the RIS-BS channel $\bm{h}^A_k$, and the transmit signal $s_k$.

\subsection{Localization Algorithm}

From (P1'), it is evident that we cannot directly estimate the user's location since the scatter locations are unknown. Therefore, we need to estimate the locations of the scatters. However, considering that scattering paths are typically weaker than direct user-RIS path, it is difficult to estimate the scatter locations directly from the received signals. To tackle this issue, we leverage the idea of successive interference cancellation (SIC) that first estimates the stronger direct user-RIS paths, and then estimates the weaker scattering paths.

Therefore, we devise a three-step algorithm to solve (P1'). Specifically, we first ignore the scattering path and roughly estimate the locations of the users based on the received signal. The residual signal is then calculated by subtracting the direct user-RIS path component from the received signal to improve the estimation accuracy of scatter locations. Next, the scatter locations are estimated using the residual signal. In the third step, we subtract the scattering path from the received signal and re-estimate the user's location to achieve higher user localization accuracy. In the following, we introduce these subproblems and the algorithm in detail. The proposed algorithm is summarized in \textbf{Algorithm~\ref{alg:1}}.

\begin{algorithm}[t]
	\caption{The proposed localization algorithm}
	\label{alg:1}
	\begin{algorithmic}[1]
		\REQUIRE{Recieved signal $\bm{G}^{(c)}$, the atom channels matrix $\bm{F}$, max number of scatters $L_{max}$} 
		\ENSURE{The estimated locations of the users $\bm{P}^{(c)}$}
		\STATE Initialization: the atom signals $\bm{\Lambda}^k$, the received signals energy $E$, the support set $\mathcal{X} = \emptyset$.
		\FOR{$k = 1,...,K$}
			\STATE Obtain the coarse estimated user locations based on (\ref{equ:19})
			\STATE Calculate residual signal by (\ref{equ19}) and (\ref{equ20})
		\ENDFOR
		\STATE Initialize the residual signal $\bm{e}_k^{ini} = \bm{e}_k$
		\FOR{$l = 1,...,L_{max}$}
			\STATE Obtain the estimated $l$-th scatter location by (\ref{equ22})
			\STATE Update support set $\mathcal{X} = \mathcal{X} \cup i$
			\STATE Calculate residual signals by (\ref{equ23}) and (\ref{equ25})
			\STATE Calculate the energy of the residual signal $E^{r}$
			\STATE \textbf{if} $E^{r} < \gamma E$ \textbf{then} break  \textbf{end if}
		\ENDFOR
		\FOR{$k = 1,...,K$}
			\STATE Calculate the new residual signal by (\ref{equ29})
			\STATE Obtain the estimated location of $k$-th user by (\ref{equ30})
		\ENDFOR
		\STATE \textbf{Return:} $\bm{P}^{(c)} = [\bm{p}_1^{(c)},...,\bm{p}_K^{(c)}]$
	\end{algorithmic}  
\end{algorithm}

\subsubsection{User Location Estimation} 
In this part, we estimate the location of each user using the corresponding received signal. This is a coarse estimate because we ignore the scattering path in this step. Specifically, we find the atom signal with the highest energy contribution to the received signals and estimate the gain of the atom channel~\cite{b38}~\cite{b37}. For example, for the $k$-th user, we use the $\bm{g}_k$ to estimate its location, i.e.,
\begin{subequations}
    \label{suba}
	\begin{align}
		\text{P1-a}: \mathop{\min}_{\bm{u}^{\bm{P}}_k,\forall k}  &\sum_{k=1}^{K} \left\Vert \bm{g}^{(c)}_k - \bm{\Lambda}^k \bm{u}^{\bm{P}}_k  \right\Vert_2^2 ,\label{suba-1} \\
		s.t. \ & \left\Vert \bm{u}^{\bm{P}}_k \right\Vert_0 = 1, \forall k.
	\end{align}	
\end{subequations}
Only one element in $\bm{u}^{\bm{P}}_k$ is nonzero, therefore $\bm{u}^{\bm{P}}_k$ is a $1$-sparse vector, which enables the sparse recovery method to efficiently recover it\cite{b40}. Note that the terms of summation in (\ref{suba-1}) are independent, therefore (P1-a) can be decoupled into $K$ independent subproblems, given by
\begin{subequations}
	\begin{align}
		\text{P1-a-k}: \mathop{\min}_{\bm{u}^{\bm{P}}_k}  & \left\Vert \bm{g}^{(c)}_k - \bm{\Lambda}^k \bm{u}^{\bm{P}}_k  \right\Vert_2^2 , \\
		s.t. \ & \left\Vert \bm{u}^{\bm{P}}_k \right\Vert_0 = 1.
	\end{align}	
\end{subequations}

The received signal $\bm{g}^{(c)}_k$ contains multiple paths. Since the gain of the direct path is larger than that of the scattering paths, we choose the atom signal with the strongest correlation with received signal as the direct path corresponding to the user's location. Mathematically, for the $k$-th user, we have
% Due to the greater path loss and the reflection of the scatter, the path gain of the direct user-RIS path is higher than those of the user-scatter-RIS paths. Hence, we select the location corresponding to the atom signal that best matches the $k$-th received signal as the estimated location for the $k$-th user. 
\begin{align}
	\label{equ18}
	i^k = \arg \max_i \left\vert (\bm{\Lambda}^k_i)^H \bm{g}^{(c)}_k \right\vert 
\end{align}
where $\bm{\Lambda}^k_i \in \mathbb{C}^{N\times 1}$ is the $i$-th column of $\bm{\Lambda}^k$ that corresponds to $\bm{z}_i$, the $i$-th location in $\bm{Z}$. Then, we have $\hat{\bm{p}}_k = \bm{z}_{i^k}$, and the estimated location for all $K$ users is given by
\begin{align}
    \label{equ:19}
    \hat{\bm{P}} = \left[\bm{z}_{i^1},...,\bm{z}_{i^K}\right].
\end{align}

Note that the optimal gain for any given atom signal 
% Mathematically, we first calculate the correlation matrix between the received signal and the atom signals, given by
% \begin{align}
% 	\bm{\Gamma} = \bm{\Lambda}_k^H \bm{g}^{(c)}_k.
% \end{align}
% Then the location corresponding to the atom signal with the highest power is selected as the location of the $k$-th user.
% \begin{align}
% 	i^{\star} = \mathop{\arg\max}_i \vert \Gamma_{i} \vert^2,
% \end{align}
% The gain of the atom signal, i.e. the estimated value of the $(i^{\star},k)$-th element in $\bm{U}^{\bm{P}}$, can be calculated by
% \begin{align}
% 	\label{equ20}
% 	\hat{u}_{i^{\star},k}^{\bm{P}} = \bm{\Lambda}_{i^{\star}}^{\dag} \bm{g}^{(c)}_k,
% \end{align}
% where $\bm{\Lambda}_i$ is the $i$-th column of the atom signals $\bm{\Lambda}$. $^{\dag}$~represents the pseudo-inverse. The residual signal for the $k$-th user can be given by
% \begin{align}
% 	\bm{E}_k = \bm{g}^{(c)}_k - \hat{u}_{i^{\star}}^{\bm{P}} \bm{\Lambda}_{i^{\star}} .
% \end{align}	
can be given in closed form as a function $i$
\begin{align}
	\label{equ19}
	\hat{u}_{ik}^{\bm{P}} ={\bm{\Lambda}^k_i}^{\dag} \bm{g}^{(c)}_k,
\end{align}
where $\hat{u}_{ik}^{\bm{P}}$ is the estimated value of the $i$-th element in $\bm{u}^{\bm{P}}_k$, and $^{\dag}$~represents the pseudo-inverse. 
% By inserting (\ref{equ20}) into (\ref{equ19}), the problem is converted to a exhaustive search
% \begin{align}
% 	\label{equ27}
% 	i =  \mathop{\arg\min} \Vert \bm{\Pi}_{\bm{\Lambda}^k_i}^{\bot} \bm{g}^{(c)}_k \Vert_2^2,
% \end{align}
% where $\bm{\Pi}_{\bm{X}}^{\bot} = \bm{I} - \bm{\Pi}_{\bm{X}}$, and $\bm{\Pi}_{\bm{X}} = \bm{X}\bm{X}^{\dag}$ denotes the orthogonal projection.
The residual signal for the $k$-th user can be given by
\begin{align}
	\label{equ20}
	\bm{e}^{(c)}_k = \bm{g}^{(c)}_k -\hat{u}_{ik}^{\bm{P}} \bm{\Lambda}^k_i.
\end{align}
In this way, signals with smaller interference for scatter localization is obtained. 

\subsubsection{Scatter Location Estimation}
In this part, the locations of the scatters are estimated using all the residual signals. This part is iteratively conducted until the energy of the new residual signal is less than a percentage of the energy of the received signals or the number of iterations reaches $L_{max}$. This subproblem is modeled as
\begin{subequations}
	\begin{align}
		\text{P1-b}: \mathop{\min}_{\bm{u}^{\bm{Q}}_k,\forall k}  &\sum_{k=1}^{K} \left\Vert \bm{e}^{(c)}_k - \bm{\Lambda}^k\bm{u}^{\bm{Q}}_k \right\Vert_2^2 , \\
		s.t. &\left\vert \{i\vert \bm{v}_{i} \neq \bm{0}\} \right\vert \leq L_{max},
	\end{align}	
\end{subequations}%
where $\bm{v}_{i} = [u_{i1}^{\bm{Q}},...,u_{iK}^{\bm{Q}}] \in \mathbb{C}^{1 \times K}$. Since the numbers of scatters are generally much smaller than the number of atom channels\cite{b54}, we can use the sparse recovery method\cite{b38} to estimate the sparse vector $\bm{u}^{\bm{Q}}_k$\cite{b40}. We define the support set $\mathcal{X}$ as the indices of the atom signals corresponding to the estimated scatter locations. In each iteration, a new index is added to the support set. Note that the atom signal for each user is different, the residual signal for different users is approximated using their corresponding atom signals, and then the location with the maximum correlation with the residual signal is selected as the estimated location of a scatter. Mathematically, we have
\begin{align}
	\label{equ22}
	i = \arg \max_i \sum_{k=1}^K \left\vert (\bm{\Lambda}^k_i)^H \bm{e}^{(c)}_k \right\vert 
\end{align}

Then, we update the support set as $\mathcal{X} = \mathcal{X} \cup i$. The gains of all the support signals in the support set are calculated through orthogonal least square, which is given by
\begin{align}
	\label{equ23}
	\hat{\bm{u}}^{\bm{Q}}_{k,\mathcal{X}} = ({\bm{\Lambda}^{k}_\mathcal{X}})^{\dag} \bm{e}_k^{ini},
\end{align}
where $\bm{e}_k^{ini}$ is the initial residual signal, and $\bm{\Lambda}_{\mathcal{X}}^k$ is the matrix generated by the atom signals selected by the indices in the support set, which is given by
\begin{align}
	\bm{\Lambda}^{k}_\mathcal{X} = [\bm{\Lambda}^{k}_{j}], j \in \mathcal{X}.
\end{align}
Using the estimated gain, the residual signal is updated by 
\begin{align}
	\label{equ25}
	\bm{e}^{(c)}_k = \bm{e}_k^{ini} - \bm{\Lambda}^{k}_\mathcal{X} \hat{\bm{u}}^{\bm{Q}}_{k,\mathcal{X}}.
\end{align}

Then the energy of the new residual signal $\bm{e}_k$ is calculated and denoted by $E^{r}$. If $E^{r} < \gamma E$, the iteration terminates, else the steps (\ref{equ22}) to (\ref{equ25}) are repeated, where $\gamma$ is a predetermined parameter and $E$ is the energy of the original received signals.

\subsubsection{User Location Refinement}
In this part, we optimize the locations of users based on the estimated scatter locations, i.e., re-solve the locations of users given the locations of scatters. First, we calculate the residual signal $\bm{r}^{(c)}_k$ by subtracting the scattering path from the original received signal.
\begin{align}
	\label{equ29}
	\bm{r}^{(c)}_k = \bm{g}^{(c)}_k - \bm{\Lambda}^{k}_\mathcal{X}{\bm{\Lambda}^{k}_\mathcal{X}}^{\dag} \bm{g}^{(c)}_k.
\end{align}
Then similar to the method of the user location estimation step, by solving the following subproblem,
\begin{subequations}
	\begin{align}
		\text{P1-c}: \mathop{\min}_{\bm{u}^{\bm{P}}_k,\forall k}  &\sum_{k=1}^{K} \left\Vert \bm{r}^{(c)}_k -\bm{\Lambda}^k\bm{u}^{\bm{P}}_k \right\Vert_2^2 , \\
		s.t. \ & \left\Vert \bm{u}^{\bm{P}}_k \right\Vert_0 = 1, \forall k,
	\end{align}	
\end{subequations}
the index of the refined $k$-th user location is given by
\begin{align}
	\label{equ30}
	i^k = \arg \max_i \left\vert (\bm{\Lambda}^k_i)^H \bm{r}^{(c)}_k \right\vert. 
\end{align}
Then we have $\hat{\bm{p}}_k = \bm{z}_{i^k}$ and $\hat{\bm{P}} = [\bm{z}_{i^1},...,\bm{z}_{i^K}]$. 
% Then we repeat the process of (18)-(21) by replacing $\bm{g}^{(c)}_k$ with $\bm{E}_k$.

% The iteration stops when the estimated locations of the users converge, i.e. the estimated $\bm{U}^{\bm{P}}$ is the same with the last iteration. Then the locations of the users can be determined by the non-zero element in $\bm{U}^{\bm{P}}$.
% In this part, we determine which elements corresponds to the direct user-RIS paths. Due to the greater path loss and the reflection of the scatter, the path gain of the direct user-RIS path is higher than those of the user-scatter-RIS paths. Hence, we select the element with the largest absolute value in each column of $\hat{\bm{U}}$ as the index of the non-zero element for the direct user-RIS path component, which can be written as
% \begin{align}
% 	i_k = \mathop{\arg\max}_i \vert \hat{{u}}_{ik}\vert,
% \end{align}
% where $\hat{{u}}_{ik}$ is the $(i,k)$-th element in $\hat{\bm{U}}$. The ${(i_k,k)}$-th element in the estimated direct path component $\hat{\bm{U}}_{\bm{P}}$ can be given by
% \begin{align}
% 	\hat{{u}}_{\bm{P}{i_k,k}} = \hat{{u}}_{i_k,k},
% \end{align}
% and the rest elements of $\hat{\bm{U}}_{\bm{P}}$ equal $0$.
% Then the estimated location of the $k$-th user is $\bm{p}_k^{(c)} = \bm{z}_{i_k}$.

\section{RIS Phase Shift Optimization}
\label{section:prob}
In this section, we first formulate the RIS phase shift optimization problem and propose an algorithm to solve it.

\subsection{Phase Shift Optimization Problem Formulation}
The selected optimization metric is CRB, a standard metric for assessing the parameter estimation error\cite{b79}. The CRB is derived from the model of the received signal for $c + 1$ cycles, which is given by
\begin{align}
	\label{rsm}
	\bm{y}_k =[y^{(1)}_k,...,y^{(c+1)}_k]^T=(\bm{B}^{(c+1)})^T\bm{h}_ks+\bm{\epsilon},
\end{align}
where $\bm{\epsilon}=[\epsilon^{(1)},...,\epsilon^{(c+1)}]^T$ is the independent and identically distributed zero-mean Gaussian noise with variance $\sigma^2$. Note that the received signal is influenced by the RIS phase shift $\bm{\beta}^{(c+1)}$, which is the $(c+1)$-th column of $\bm{B}^{(c+1)}$. Thus, we can obtain the expressions between $\bm{\beta}^{(c+1)}$ and CRB, listed in the following. By optimizing the RIS phase shifts $\bm{\beta}^{(c+1)}$, we can minimize the CRB to improve the localization accuracy.

Specifically, we formulate the optimization problem to minimize the sum of weighted CRBs of range and angle estimation errors. We introduce a weight matrix to address the different unit of angle and range CRBs~\cite{b88}. Note that we estimate angles and ranges for NF users, while only angles for FF users. Hence, we give the expressions of the weighted CRBs for NF and FF users separately.

\subsubsection{CRBs for NF Users}
If the estimated location $\bm{p}_k^{(c)}$ is in the NF region in the $c$-th cycle, we expect that in the $(c+1)$-th cycle, the range and angles of the $k$-th user are estimated, denoted by $\bm{p}_k^{(c+1)}=[R_k,\theta_k, \phi_k]^T$. The expressions of CRBs for the NF case are provided in proposition~\ref{prop:NFCRB}.

\begin{proposition}
	\label{prop:NFCRB}
	For the NF case, the CRBs of the unknown parameters $\bm{p}_k = [R_k,\theta_k,\phi_k]^T$ are given by
	{
		\begin{align}
			\label{equation:nf_crb}
			D_R &=\left(\bm{J}_{NF}^{-1}\right)_{1,1},\\
			D_\theta &=\left(\bm{J}_{NF}^{-1}\right)_{2,2},\\
			D_\phi &=\left(\bm{J}_{NF}^{-1}\right)_{3,3},
		\end{align}
	}%
	where $\bm{J}_{NF}$ is the $3 \times 3$ Fisher information matrix of $\bm{p}_k$ in the NF region, and the $(i, j)$-th element in $\bm{J}_{NF}$ is given by~\cite{b74}
	\begin{equation}
		\label{prop2}
		[\bm{J}_{NF}]_{i,j} = \dfrac{2}{\sigma^2} \sum_{m=1}^{c+1}Re \left \{ \dfrac{\partial (\mu^{(m)})^H}{\partial p_i} \dfrac{\partial \mu^{(m)}}{\partial p_j} \right \},
	\end{equation}
	where $\mu^{(m)}$ is defined as the noise-free received signal
	\begin{align}
        \label{equmu1}
		\mu^{(m)}=(\bm{\beta}^{(m)})^T \bm{h}^{NF}_k s_k.
	\end{align} 
	Here $\bm{h}^{NF}_k$ is the channel calculated by using $\bm{p}_k^{(c)}\in \bm{D}_{NF}$ and $\bm{Q}^{(c)}$, and $p_i$ is the $i$-th element in $\bm{p}_k$. 

	\begin{IEEEproof}
		See Appendix \ref{appen:NFCRB}.
	\end{IEEEproof}
	
\end{proposition}

\subsubsection{CRBs for FF Users}
If the estimated location $\bm{p}_k^{(c)}$ is in the FF region, we expect that only the angles are estimated for the $k$-th user in the $(c+1)$-th cycle, and the estimated parameters are denoted by $\bm{p}_k^{(c+1)}=[\theta_k, \phi_k]^T$. The expressions of CRBs for the FF case are provided in proposition~\ref{prop:FFCRB}.

\begin{proposition}
	\label{prop:FFCRB}
	For the FF case, the CRBs of the unknown parameters $\bm{p}_k$ are given by
	\begin{align}
		D_\theta &=\left(\bm{J}_{FF}^{-1}\right)_{1,1},\\
		D_\phi &=\left(\bm{J}_{FF}^{-1}\right)_{2,2},
	\end{align}
	where $\bm{J}_{FF}$ is the $2\times 2$ Fisher information matrix of $\bm{p}_k$ in the FF region, and the $(i, j)$-th element is given by~\cite{b74}
	\begin{equation}
		\label{prop3}
		[\bm{J}_{FF}]_{i,j} = \dfrac{2}{\sigma^2} \sum_{m=1}^{c+1}Re \left \{ \dfrac{\partial (\mu^{(m)})^H}{\partial p_i} \dfrac{\partial \mu^{(m)}}{\partial p_j} \right \},
	\end{equation}
	where $\mu^{(m)}$ is defined as the noise-free received signal
	\begin{align}
        \label{equmu2}
		\mu^{(m)}=(\bm{\beta}^{(m)})^T \bm{h}^{FF}_k s_k,
	\end{align}
	and $\bm{h}^{FF}_k$ is the channel calculated using $\bm{p}_k^{(c)}$ and $\bm{Q}^{(c)}$.

	\begin{IEEEproof}
		See Appendix \ref{appen:FFCRB}.
	\end{IEEEproof}
\end{proposition}

\subsubsection{Optimization Problem Formulation}
To achieve high localization accuracy for multiple users, we optimize the phase shifts $\bm{\beta}^{(c+1)}$ by minimizing the sum of CRBs of all the users. The optimization problem can be formulated as
\begin{subequations}
	\begin{align}
		\text{P2}: \mathop{\min}_{\bm{\beta}^{(c+1)}} f(\bm{P},\bm{\beta}^{(c+1)}) &= \sum_{k \in \Psi_{NF}} tr \left(\bm{J}_{NF,k}^{-1}\bm{W}_{NF} \right) \nonumber \\ 
		& + \sum_{k \in \Psi_{FF} } tr \left(\bm{J}_{FF,k}^{-1}\bm{W}_{FF} \right),\\
		 s.t. \ \vert&\beta_{n}^{(c+1)} \vert=1,\ \forall n=1,2,...N, \label{eq:sub27b}
	\end{align}
\end{subequations}
where $\bm{J}_{NF,k}$ or $\bm{J}_{FF,k}$ is the Fisher information matrix of $\bm{p}_k$ when the $k$-th user is located in the NF or FF region and is a function of the RIS phase shifts $\bm{\beta}^{(c+1)}$. Here $\bm{W}_{NF}=\text{diag} \{w_1,w_2,w_3 \}$ and $\bm{W}_{FF}=\text{diag} \{w_4,w_5 \}$ are the weight matrices of the CRBs in the NF and FF regions, respectively~\cite{b88}. $\Psi_{NF}$ and $\Psi_{FF}$ are the sets of estimated NF and FF users, which are mathematically given by
\begin{align}
	\Psi_{NF} = \{k|\bm{p}_k^{(c)} \in \bm{D}_{NF} \}, \\
	\Psi_{FF} = \{k|\bm{p}_k^{(c)} \in \bm{D}_{FF} \}.
\end{align}
$\beta_{n}^{(c+1)}$ is the RIS phase shift of the $n$-th element in the $(c+1)$-th cycle, satisfying the constant modulus constraint (\ref{eq:sub27b}). 

% Note that unlike single-user localization, for multi-user joint localization, all users share the same RIS phase shifts. Therefore, it is necessary to design a RIS phase shift optimization method to maximize the localization accuracy of multiple users simultaneously.

\subsection{RIS Phase Shift Optimization Algorithm}

We design an optimization algorithm based on the complex circle manifold~(CCM) method to tackle (P2). Due to the constant modulus constraint (\ref{eq:sub27b}), the problem (P2) is non-convex, which is numerically difficult to handle. Fortunately, the solution can be considered as lying on the CCM to satisfy the constant modulus, where the manifold is represented as
\begin{flalign}
    \mathcal{M}^N= \! \left\{ \bm{\beta}^{(c+1)} \in \mathbb{C}^{N} \! :|\beta_1^{(c+1)}|=... \! =|\beta_N^{(c+1)}|=1 \! \right\}.
\end{flalign}

The main idea of the CCM-based optimization method is to iteratively apply gradient descent in the manifold space. After several iterations, the algorithm terminates when the difference between two iterations of $f(\bm{P},\bm{\beta}^{(c+1)})$ is less than a constant $\zeta$ or the number of iterations exceeds $I$, where $\zeta$ and $I$ are selected to comply with the time constraints of the optimization step. The CCM-based optimization method is composed of four main steps in each iteration:

\subsubsection{Compute the gradient in Euclidean space}
We use the Euclidean gradient as the search direction for the minimization problem in Euclidean space. The Euclidean gradient of $f(\bm{\beta}^{(c+1)})$ is given by
\begin{align}
\label{eq:nabla}
    \nabla f(\bm{\beta}^{(c+1)}) \!
    = \! 2 \Big( \! &\sum_{k \in \Psi_{NF}} \! \left[ \dfrac{ w_1 \partial D_{R,k}}{\partial{\bm{\beta}^{(c+1)}}^{\ast}} + \! \dfrac{w_2 \partial D_{\theta,k}}{\partial{\bm{\beta}^{(c+1)}}^{\ast}} + \! \dfrac{w_3 \partial D_{\phi,k}}{\partial{\bm{\beta}^{(c+1)}}^{\ast}} \right] \nonumber \\
    + &\sum_{k \in \Psi_{FF}}\left[ \dfrac{w_4 \partial D_{\theta,k}}{\partial{\bm{\beta}^{(c+1)}}^{\ast}} +  \dfrac{w_5 \partial D_{\phi,k}}{\partial{\bm{\beta}^{(c+1)}}^{\ast}} \right] \Big),
\end{align}
where the variable $\bm{P}$ is omitted in $f(\bm{P},\bm{\beta}^{(c+1)})$ for simplicity. Similar to the proof in \cite{b14}, the specific expressions of the differentials of CRBs are given in Appendix \ref{appen:dCRB}.

\subsubsection{Compute the Riemannian gradient}
The Riemannian gradient is the projection of the Euclidean gradient onto the tangent space of the complex circle manifold. The Riemannian gradient of the objective function $f(\bm{\beta}^{(c+1)})$ at the point $\bm{\beta}^{(c+1)}_j$ on the complex circle manifold $\mathcal{M}$ is given as~\cite{b14}
\begin{align}
	\label{eq:nfg}	\nabla_{\mathcal{M}}&f(\bm{\beta}^{(c+1)}_j)= -\nabla f(\bm{\beta}^{(c+1)}_j) \nonumber \\
	&-Re \left\{ \left(\nabla f(\bm{\beta}^{(c+1)}_j)\right)^{*} \odot {\bm{\beta}^{(c+1)}_j} \right\} \odot \bm{\beta}^{(c+1)}_j.
\end{align}

\subsubsection{Update over the Tangent Space}
We choose a step size to update the current point, which is mathematically given as
\begin{align}
\label{equation:barj}
	\bar{\bm{\beta}}^{(c+1)}_j = \bm{\beta}^{(c+1)}_j  + \alpha_j \nabla_{\mathcal{M}}g(\bm{\beta}^{(c+1)}_j),
\end{align}
where $\alpha_j$ is the step size in the $j$-th iteration.

\subsubsection{Retract onto the manifold}
After the update, the new point $\bar{\bm{\beta}}^{(c+1)}_j$ generally does not lie on the manifold $\mathcal{M}$. By using the retraction operator, the new point is mapped into the manifold. The retraction operator is given as
\begin{align}
\label{equation:j1}
	\bm{\beta}^{(c+1)}_{j+1}= \bar{\bm{\beta}}^{(c+1)}_j \odot \dfrac{1}{ |\bar{\bm{\beta}}^{(c+1)}_j |}.
\end{align}

The proposed algorithm is summarized in \textbf{Algorithm~\ref{alg:algorithm1}}.

\begin{algorithm}[t]
	\caption{CCM-based RIS Phase Shift Optimization Algorithm}
	\label{alg:algorithm1}
	\begin{algorithmic}[1]
		\REQUIRE{Estimated location $\hat{\bm{p}}^{(c)}$, RIS phase shift for previous $c$ cycles $\bm{B}^{(c)}$}
		\ENSURE{Optimal RIS phase shift $\bm{\beta}^{(c+1)}$}  
		\STATE Initialize: $j=0$, $\beta_0 \in \mathcal{M}$;
		\WHILE{$|g(\bm{\beta}^{(c+1)}_{j+1}) - g(\bm{\beta}^{(c+1)}_{j})| > \zeta $ and $j < I$} {
		\STATE Compute the Euclidean gradient $\nabla f(\bm{\beta}^{(c+1)}_j)$ according to (\ref{eq:nabla}) }
		\STATE Calculate the Riemannian gradient $\nabla_{\mathcal{M}}f(\bm{\beta}^{(c+1)}_j)$ according to (\ref{eq:nfg}); 
		\STATE Compute the RIS phase shift update on the tangent space $\bar{\bm{\beta}}^{(k+1)}_j$ according to (\ref{equation:barj}); 
		\STATE Update RIS phase shift $\bm{\beta}^{(k+1)}_{j+1}$ according to (\ref{equation:j1}); 
		\STATE $j = j+1$;
		\ENDWHILE
	\end{algorithmic}
\end{algorithm}

\section{Performance Analysis}
\label{section:peran}
In this section, we analyze the complexity of the proposed method and discuss its localization performance.
\subsection{Algorithm Complexity}
\subsubsection{Complexity of the localization algorithm}
% \subsubsection{Complexity of the coarse estimation phase}
For the user location estimation step of the localization algorithm, the method is similar with the first iteration of the OMP method. Therefore, the computational cost of the user estimation step is $\mathcal{O}(cMK)$\cite{b69}.
% note that (\ref{equ19}) is equivalent to the following problem\cite{b20}
% \begin{align}
% 	\hat{\bm{p}}_k = \mathop{\arg\max}_{\bm{p}_i \in \bm{Z}} \dfrac{\bm{\Lambda}_k(\bm{p}_i)^H \bm{g}^{(c)}_k}{\Vert \bm{\Lambda}_k(\bm{p}_i) \Vert^2}.
% \end{align}
% Let us define $\bm{x}_i = \text{diag}\{\bm{h}^A_k\}\bm{F}_i$. In the $c$-th cycle, the computational cost of $\bm{\Lambda}_k(\bm{p}_i) = (\bm{B}^{(c)})^T \bm{x}_i s_k$ is equal to $\mathcal{O}(cN)$. Then we need to search over all candidate locations, i.e. $M = S + N_{\theta}N_{\phi}$ possible locations for $K$ users. Therefore, the cost of the user estimation step is $\mathcal{O}(cNMK)$.
% \subsubsection{Complexity of the location refinement phase}
In the scatter location estimation step, the complexity of this algorithm has been analyzed in \cite{b71}. We estimate the scatter locations for at most $L_{max}$ rounds. Hence, the computational cost is $\mathcal{O}(cKML_{max})$. Then, similar with the user location estimation step, the computational cost of user location refinement step is $\mathcal{O}(cMK)$.

Hence, the overall cost of the localization algorithm for the $c$-th cycle is $\mathcal{O}(cMK) + \mathcal{O}(cKML_{max})$. The overall complexity for a $C$-cycle localization algorithm is $\mathcal{O}(C(C+1)MK) + \mathcal{O}(C(C+1)KML_{max})$.

\subsubsection{Complexity of the RIS optimization algorithm}
The complexity of the CCM algorithm has been analyzed in \cite{b41}. 
% Since the process of finding the range of hyperparameters to guarantee convergence can be done offline, the complexity of the optimization algorithm is solely determined by the iteration process. 
Let us denote the total number of iterations required to converge by $T_{CCM}$. Then the total complexity of the optimization algorithm is $\mathcal{O}(T_{CCM}N^2)$. Since the optimization algorithm is conducted for $C-1$ times, the overall complexity for the optimization algorithm is $\mathcal{O}((C-1)T_{CCM}N^2)$.

\subsection{Localization Performance}

% In order to reflect both the range and angle estimation error, we first select the midpoint of the FF region as the representative range for FF locations, and the new candidate locations are given by $\bm{Z}'$. Then the error parameter is defined as

% To simplify the analysis, we ignored the scattering path. Hence this analysis result of the localization error is the lower bound of the actual error. The reason for doing this is twofold: firstly, the scattering path is generally weak; secondly, given appropriate sampling methods and RIS phase shifts selection, the correlation between the direct path and the scattering path is limited, resulting a similar accuracy with no scattering path in the refinement stage. The second point is shown in the literature. According to \cite{b42}, the FF steering vector has asymptotic orthogonality, and based on \cite{b37}, by selecting appropriate NF range sampling spacing, the correlation of NF steering vectors could be sufficient low.

% To simplify the analysis, we ignored the scattering path. Hence this analysis result of the localization error is the lower bound of the actual error. The reason for doing this is twofold: firstly, the scattering path is generally weak; secondly, given appropriate RIS phase shift selection, the correlation between the direct path and the scattering path could be sufficiently low, resulting in a similar accuracy with no scattering path in the refinement stage.

Localization errors primarily arise from two sources: one is the failure to estimate the grid closest to the user, which we refer to as grid misjudgment; the other is the error between the user's continuous location in actual space and the discrete grids, known as grid mismatch, reflecting the inherent limitation of discretizing continuous space\cite{b45}. Mathematically, suppose the real location of the $k$-th user is $\bm{p}_k$, the estimation result is $\bm{z}_{m'}$, and the nearest grid for the $k$-th user is $\bm{z}_m$. Then the average localization error can be given by
\begin{align}
	\mathbb{E}(l_e) &= \mathbb{E}\{\Vert \bm{p}_k - \bm{z}_{m'} \Vert\} \nonumber \\
	&\leq \mathbb{E}\{\Vert \bm{p}_k - \bm{z}_m \Vert\} + \mathbb{E}\{\Vert \bm{z}_m - \bm{z}_{m'} \Vert\},
\end{align}
where the first term corresponds to grid mismatch, and the second term is grid misjudgment. In the following, we analyze these two sources of errors respectively.
% $l_e = \Vert \bm{p}_k - \bm{z}_{m'} \Vert$ is the localization error.

\subsubsection{Number of Cycles}
The expectation of localization error can be given by\cite{b30}
\begin{flalign}
	\label{eq:error}
	\mathbb{E}(l_e) \! = \sum_{m=1}^{M} \gamma_{m}^{k} \! \int_{\bm{g}_k^{(C)} \! \in \mathcal{R}_{km}} \! \!\!\mathbb{P}(\bm{g}_k^{(C)} | \bm{B}^{(C)}, \bm{p}_k, \bm{Q}) d\bm{g}_k^{(C)}\!, \!\!\! 
\end{flalign}
where $\mathcal{R}_{km}$ is the decision region for the $m$-th candidate location. The integral in (\ref{eq:error}) is the probability that the estimated location is the $m$-th candidate location, given the ground truth that the location of the $k$-th user is $\bm{p}_k$. Here $\gamma_{m}^{k}$ is the error parameter, defined as $\gamma_{m}^{k} = \Vert \bm{p}_k - \bm{z}_m \Vert$.

We assume all locations have the same prior probabilities. Hence, the decision region can be given by
\begin{align}
	\mathcal{R}_{km} = \{ \bm{g}_k^{(C)}: &\mathbb{P}(\bm{g}_k^{(C)} | \bm{B}^{(C)}, \bm{z}_m, \bm{Q}) \nonumber\\ 
	&\geq \mathbb{P}(\bm{g}_k^{(C)} | \bm{B}^{(C)}, \bm{z}_{m'}, \bm{Q}), \forall m' \neq m\}.
\end{align}
Since the noise obeys Gaussian distribution, we have
\begin{flalign}
    \mathcal{R}_{km} \! = \! \{ \bm{g}_k^{(C)} \! : \! \vert \bm{g}_k^{(C)} \! -\! \bm{\mu}_m \vert^2 \! \leq \! \vert \bm{g}_k^{(C)} \! - \!\bm{\mu}_{m'} \vert^2 \!,  \forall m' \! \neq \! m \},
\end{flalign}
where $\bm{\mu}_m = [\mu^{(1)}_m,...,\mu^{(C)}_m]^T$ is the noise-free signal for the $m$-th grid. Let $\bm{\xi}_k = \bm{g}_k^{(C)} - \bm{\mu}_m $, we have\cite{b30}
\begin{align}
	\label{eq:R}
	\mathcal{R}_{km} = &\{ \bm{g}_k^{(C)}: \vert \bm{\xi}_k \vert^2 \leq \vert \bm{\xi}_k + \bm{\mu}_m - \bm{\mu}_{m'}\vert^2,\forall m' \neq m  \}, \nonumber \\
	% =&\{ \bm{g}_k^{(C)}: \vert \bm{\xi}_k \vert^2 \leq \vert \bm{\xi}_k \vert^2 + \vert \bm{\mu}_m - \bm{\mu}_{m'}\vert^2 \nonumber \\ 
	% & + 2\bm{\xi}_k  (\overline{\bm{\mu}_m - \bm{\mu}_{m'}}),\forall m' \neq m \}, \\
	=&\{ \bm{g}_k^{(C)}: \vert \bm{\mu}_m - \bm{\mu}_{m'}\vert^2 + 2\bm{\xi}_k  (\overline{\bm{\mu}_m - \bm{\mu}_{m'}}) \geq 0, \\
	& \  \forall m' \neq m \nonumber\}.
\end{align}
Assuming the user is located at a grid location $\bm{z}_m$, when $C \to \infty$, the first term is greater than 0, and the second term converges to $0$ because $\bm{\xi}_k$ and $\bm{\mu}_m - \bm{\mu}_{m'}$ are independent and $\mathbb{E}(\bm{\xi}_k) = 0$. Therefore, we have
\begin{align}
	\lim_{C \to \infty} \mathbb{P} (\bm{g}_k^{(C)} \in \mathcal{R}_{km}) = 1.
\end{align}
Hence, when ignoring grid mismatch, the grid misjudgment error $\mathbb{E}\{\Vert \bm{z}_m - \bm{z}_{m'} \Vert\}$ converges to $0$.

When the user is not at a grid, if we assume the midpoint of $\bm{z}_m$ and $\bm{z}_{m'}$ also corresponds the midpoint of $\bm{\mu}_m$ and $\bm{\mu}_{m'}$, then we have $\mathbb{E}(\bm{\xi}_k) \leq \vert \bm{\mu}_m - \bm{\mu}_{m'}\vert /2 $. Hence, the $\vert \bm{\mu}_m - \bm{\mu}_{m'}\vert^2 + 2\bm{\xi}_k  (\overline{\bm{\mu}_m - \bm{\mu}_{m'}}) \geq 0$ is still alwasy true, and we still have $\lim_{C \to \infty} \mathbb{P}(\bm{g}_k^{(C)} \in \mathcal{R}_{km}) = 1$. However, the number of cycles does not affect the grid mismatch error. Therefore, the localization error would converge to a non-zero number.

\textit{Remark 1:} By increasing the number of estimation cycles~$C$, the localization error first decreases and then remains fixed.

\subsubsection{Sampling Spacing}

First, we consider the scenario $C \to \infty$. According to the previous analysis, the localization error
\begin{align}
	\label{eq:le}
	\mathbb{E}(l_e) &\leq \mathbb{E}\{\Vert \bm{p}_k - \bm{z}_m \Vert\} + \mathbb{E}\{\Vert \bm{z}_m - \bm{z}_{m'} \Vert\}, \nonumber \\
	& \overset{C \to \infty}{\approx} \mathbb{E}\{\Vert \bm{p}_k - \bm{z}_m \Vert\},
\end{align}
where $\mathbb{E}\{\Vert \bm{p}_k - \bm{z}_m \Vert\}$ is directly related to the sampling spacing. Hence, a finer grid sampling could alleviate the grid mismatch problem\cite{b45}.

When constrained by a limited number of cycle, the approximation in (\ref{eq:le}) no longer holds. Note that $\bm{\xi}_k = \bm{g}_k^{(C)} - \bm{\mu}_m $ in (\ref{eq:R}) follows Gaussian distribution. At a given SNR, decreasing the sampling spacing raises the probability that $\vert \bm{\xi}_k \vert > \vert \bm{\mu}_m - \bm{\mu}_{m'}\vert /2$. This leads to a reduction in the probability of correctly estimating the nearest grid and an increase in the expected error between the estimated and nearest grid points, i.e. $\mathbb{E}\{\Vert \bm{z}_m - \bm{z}_{m'} \Vert\}$. Furthermore, the effectiveness of sparse recovery methods is constrained by the restricted isometry property (RIP) or the mutual coherence condition. Reducing the sampling spacing may breach this condition, leading to a decline in localization accuracy\cite{b72}.
% Note that as the sampling spacing is reduced, the average estimation error of the nearest location is also reduced, while the probability of estimating the nearest locations remains nearly $1$. 

% \textit{Remark 1:} The estimation error positively related to the sampling spacings.

% However, when the sampling spacing is further reduced, the lower bound can no longer be achieved because the SNR is not sufficiently high for extremely dense sampling spacing. The localization accuracy will gradually converge to the case of continuous grids. The impact of grid mismatch is mitigated, and the grid misjudgment becomes the primary factor. At this point, finer grid sampling does not affect the accuracy.

\subsubsection{Number of RIS Elements}
Similarly to \cite{b51}, we characterize the average received power with respect to the number of RIS elements as $N \to \infty$. We consider two different RIS phase shifts: random phase shifts and optimal phase shifts that maximize the SNR. We consider $C=1$ as an example.

\begin{proposition}
	Assume $\bm{h}_k^A \sim \mathcal{CN}(\bm{0},\rho_{A}^2\bm{I})$ and $\bm{h}_k^t \sim \mathcal{CN}(\bm{0},\rho_{t}^2\bm{I})$, the average received power holds that
	\begin{equation}
		\lim_{N \to \infty} P =
		\begin{cases}
		N \rho_{A}^2 \rho_{t}^2, & \bm{\beta} = [1,...,1]^T,\\
		N^2 \dfrac{\pi^2 \rho_{A}^2 \rho_{t}^2}{16}, & \text{max SNR phase shifts}.
		\end{cases}
	\end{equation}

	\begin{IEEEproof}
            % It can be proved in a similar method in \cite{b51}.
		The two cases are discussed as follows:
		\begin{itemize}
			\item The channel $h = (\bm{h}^A_k)^H \text{diag}\{\bm{\beta}\} \bm{h}_k^t = (\bm{h}^A_k)^H \bm{h}_k^t$. According to Lindeberg-Lévy central limit theorem, we have $h\sim \mathcal{CN}(\bm{0},N\rho_{A}^2\rho_{t}^2) $ as $N \to \infty$. The average user received power is given by
			\begin{align}
				\lim_{N \to \infty}P = \mathbb{E}\vert h\vert^2 =  N \rho_{A}^2 \rho_{t}^2.
			\end{align} 
			\item For RIS phase shifts of max SNR, we have $\vert h\vert = (\bm{h}^A_k)^H \text{diag}\{\bm{\beta}\} \bm{h}_k^t = \sum_{n=1}^{N}\vert h^A_{k,n}\vert \vert h_{k,n}^t\vert$, where $h^A_{k,n}$ and $h_{k,n}^t$ are the $n$-th element in $\bm{h}^A_k$ and $\bm{h}_k^t$, respectively. Since $h^A_{k,n}$ and $h_{k,n}^t$ are statistically independent and follow Rayleigh distribution with mean values $\sqrt{\pi} \rho_{A}/2$, $\sqrt{\pi} \rho_{t}/2$, we have $\mathbb{E}(\vert h^A_{k,n}\vert \vert h_{k,n}^t\vert) = \pi \rho_{A} \rho_{t}/4$. The average user received power is given by
            \begin{flalign}
                \lim_{N \to \infty} \! P \! = \! \lim_{N \to \infty} \left\vert \sum_{n=1}^{N}\vert h^A_{k,n}\vert \vert h_{k,n}^t\vert \right\vert^2 = N^2 \dfrac{\pi^2 \rho_{A}^2 \rho_{t}^2}{16}.
            \end{flalign}
		\end{itemize}
	\end{IEEEproof}
\end{proposition}

Note that the power gain of order $N$ can be achieved for fixed phase shifts, which reveals the inherent aperture gain of a larger RIS by collecting more signal power. Moreover, setting the RIS phase shifts for max SNR can also achieve another beamforming power gain of order $N$ simultaneously.

Hence, with an increasing number of RIS elements, the SNR improves, which enlarges the difference between $\bm{\mu}_m $ and $\bm{\mu}_{m'}$. Consequently, the probability of $\vert \bm{\xi}_k \vert> \vert \bm{\mu}_m - \bm{\mu}_{m'}\vert /2$ decreases, which lowers the likelihood of grid misjudgment. However, the impact of grid mismatch persists.

\textit{Remark 2:} As the number of RIS elements increases, the localization accuracy first improves and then remains fixed. 

\begin{figure*}[t!]
	\centering
	\includegraphics[scale=0.42]{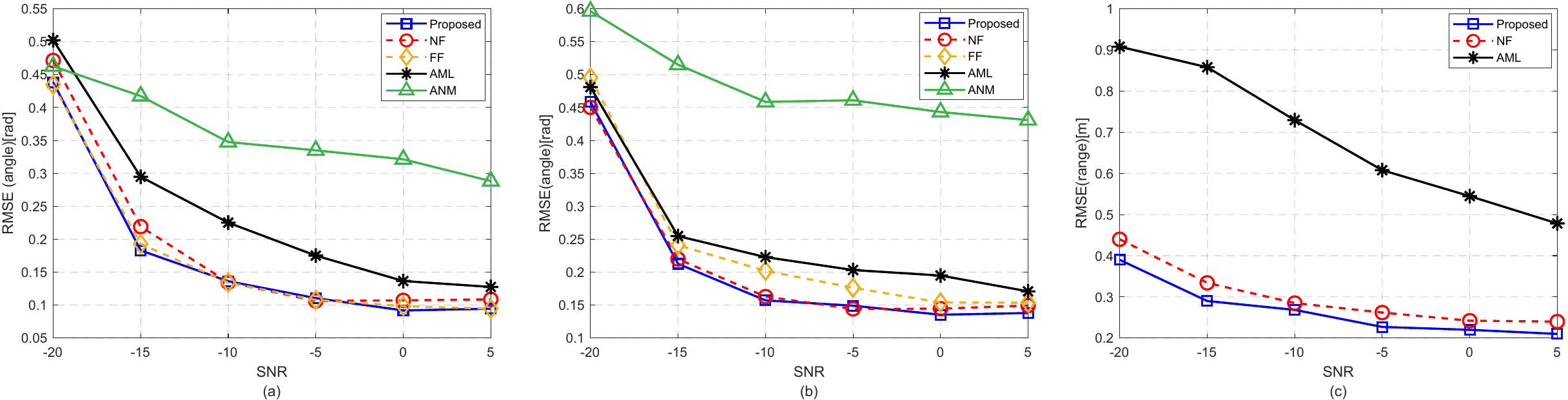}
	\caption{(a) Angle estimation accuracy of the proposed algorithm compared with AML and ANM when the user is located in the FF region. (b) Angle estimation accuracy of the proposed algorithm compared with AML and ANM when the user is located in the NF region. (c) Range estimation accuracy of the proposed algorithm compared with AML when the user is located in the NF region.}
	\label{fig:1}
\end{figure*}

\section{Simulation Results}
% Simulations are conducted on a personal computer with a 1.80GHz Intel Core i7 processor.
In this section, we present simulation results demonstrating the performance of the proposed method. The RIS is placed on the plane $x = 0$, and its center is at the origin $(0,0,0)$m. We use an RIS with $10\times 10$ elements and the element spacing is $\lambda/2 = 0.03$m. The localization range is $10$m in radius~\cite{b84}. The NF range for the considered RIS is approximately $4.9$m, while the rest region is FF. The users are equally distributed in NF and FF regions. We consider a multipath environment with $3$ randomly distributed scatters. The azimuth and elevation angles are both uniformly sampled with spacing $\pi/10$, and the range is sampled with spacing $0.25$m. The center transmit frequency is 5GHz~\cite{b87}. The number of cycles is set as $20$. We conduct $T=1000$ independent trials to obtain the average result. The simulation parameters are listed in Table~\ref{table:1} unless otherwise stated. The proposed method is compared with following methods.
\begin{itemize}
	\item Approximate maximum likelihood~(AML) scheme\cite{b20}: a RIS-enabled NF localization method based on maximum likelihood. This method decouples the angle and range parameters and iteratively searches the angle and range to fit best with the received signal.
        \item Atomic norm minimization~(ANM) scheme\cite{b53}: an off-grid direction of arrival estimation method based on atomic norm that targets FF users localization.
	\item NF scheme: The proposed algorithm is combined with pure NF model to show the gain brought by the hybrid model, labeled as NF.
        \item FF scheme: The proposed algorithm is combined with pure FF model, labeled as FF.
\end{itemize}
% and the frequency spacing between different users is $\Delta f = 120$kHz\cite{b21}

For all compared algorithms, we employ the same system setup and localization protocol. For NF methods, including AML and the proposed method with the NF model, the whole localization range is treated as in the NF, i.e., the range of FF is also sampled, while for FF methods, we only sample the angle.

\begin{table}
	\centering
	\caption{Simulation Parameters} 
	\normalsize
	\resizebox{70mm}{!}{
		\begin{tabular}{ll} 
		\toprule Parameters & Values\\ 
		\midrule 
		Number of users $K$ & $2$ \\
		Number of scatters $L$ & $3$ \\
		Transmit center frequency $f_c$ & $5$GHz \\ 
		RIS center location & $(0, 0, 0)$m \\
		Number of RIS element $N$ & $10 \times 10$ \\
		RIS element spacing & 0.005m \\
		Localization range & 10m \\
		Angle sampled spacing $\Delta \theta, \Delta \phi $ & $\pi/10$ rad \\
		Range sampled spacing $\Delta R$ i& 0.25m \\
		Number of cycles $C$ & 20 \\ 
		\bottomrule 
		\end{tabular}
	}
	\label{table:1}
\end{table}

\label{section:simu}
\subsection{Performance Evaluation}

Fig. \ref{fig:1} shows the RMSE of angle or range estimation versus the signal-to-noise ratio (SNR) for NF or FF users. The SNR is defined as $P_s / \sigma^2$, where $P_s = ((\bm{h}^A_k)^T\bm{I} \bm{h}_{k0}^t)^2$ is the received signal power of the direct path when the phase shifts of all the RIS elements are $1$. Fig. \ref{fig:1} (a) and (b) show the angle estimation RMSE of the proposed algorithm and other comparison algorithms for FF and NF users, respectively. They indicate that the proposed method outperforms all other methods for FF and NF user localization. We also observe that for algorithms other than ANM, as SNR increases, the RMSE of the angle estimation first decreases and then remains fixed. The RMSE does not decrease when the SNR is very high because the estimated location must be at one of the sampled grids, while the actual location of the users is off-grid. The ANM does not suffer from the grid mismatch problem since it is an off-grid algorithm.

Fig.~\ref{fig:1}~(a) also reveals the RMSEs of the proposed algorithm with the hybrid model and FF model are similar, indicating that the hybrid model can achieve the same accuracy for FF user localization although with the interference of NF candidate locations. We also observe that the proposed algorithm performs worse in the NF model than in the hybrid model for FF users. This is because, for the NF model, the range of the FF region is also sampled, resulting in more candidate locations in the FF region and making the algorithm more difficult to converge. At the same time, the channels of the sampled FF candidate locations in the NF model have strong correlations, which leads to energy dispersion especially when estimating the scatters locations. In Fig. \ref{fig:1} (b), we observe that the FF model has a higher RMSE than the hybrid model when estimating the NF angles. This is due to the model mismatch of the FF model when describing the NF channels. 
% , resulting in lower estimation performance. 
% We conducted more simulations in the following to support the above arguments.
% 

% Fig. \ref{fig:1} (b) shows the angle estimation RMSE of the proposed algorithm when the user is located in the NF region. 
% Furthermore, for NF user localization, the angle estimation result of the proposed algorithm applied to the hybrid and NF models is similar.
% It shows that the proposed method provides higher localization accuracy than the compared algorithms. 

Fig. \ref{fig:1} (c) shows the range estimation accuracy for NF users versus SNR. For range estimation, the proposed method outperforms AML by more than 55\% for all SNR conditions. It shows that the range estimation result of the NF model is not as good as those of the hybrid model for NF user localization, which reveals the superiority of the hybrid model. Additionally, the improvement of the hybrid model over the NF model is bigger in range compared to angle estimation. The reason for this phenomenon is that the NF model samples the range and angle in the whole region, i.e., the range of FF is also sampled. However, the steering vectors of the different ranges in the FF are very similar. If a NF user is misestimated as a FF point, the resulting range error becomes significantly larger, leading to an overall degradation in estimation accuracy.

In Fig. \ref{fig:1}, we can observe that the proposed method significantly outperforms both AML and ANM. The performance gain can be attributed to the following reasons: Firstly, neither AML nor ANM effectively handles the scattering paths, and the interference of scattering paths reduces localization accuracy. Secondly, AML decouples angle and range estimation and updates them iteratively. While this approximated method reduces algorithm complexity, it also impacts performance. Moreover, ANM is a FF localization method and employs plane wave model, which inaccurately describes NF signals.

\begin{figure}
	\centering
	\includegraphics[scale=0.5]{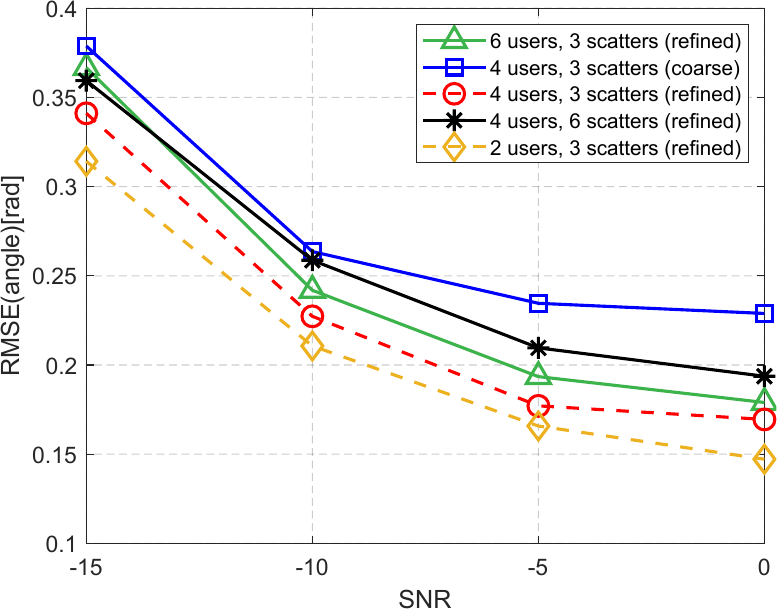}
	\caption{Angle estimation performance for different localization scenarios.}
	\label{USER}
\end{figure}

In the following, the angle sampling spacing is set as $\pi/20$ to reduce grid mismatch. Fig. \ref{USER} shows the influence of the number of users and scatters. We can observe that increasing the number of users degrades angle estimation accuracy, as the RIS beamforming becomes less precise when simultaneously serving multiple users. Moreover, increasing the number of scatters reduces localization accuracy due to interference from multipath propagation. Furthermore, by comparing the results obtained directly using the OMP algorithm, i.e., the first step of the localization algorithm (labeled as coarse) with those obtained through the three-step localization algorithm (labeled as refined), it is evident that the scattering path elimination technique can improve the localization accuracy.

\begin{figure}
	\centering
	\includegraphics[scale=0.5]{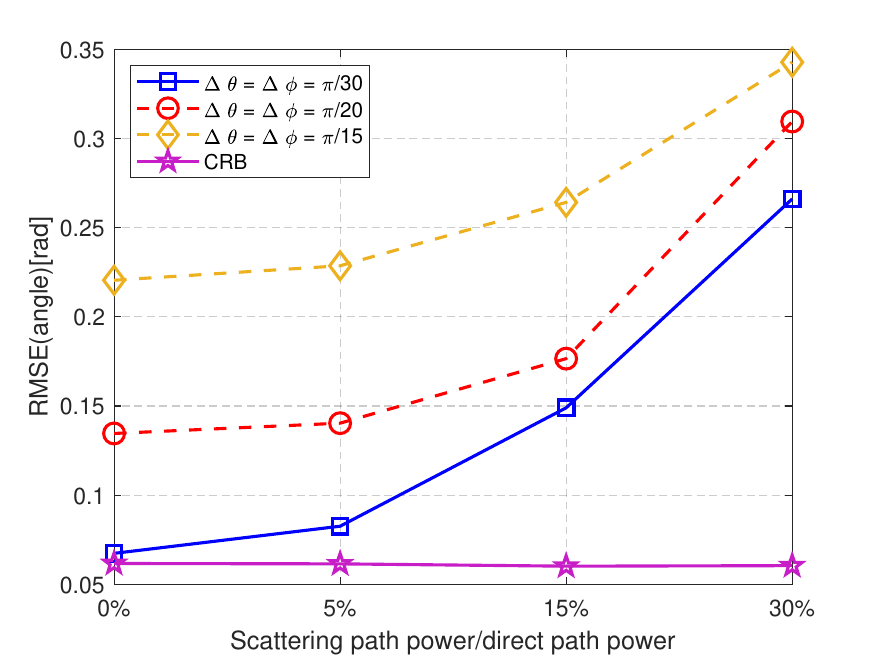}
	\caption{Angle estimation performance for different interference power.}
	\label{SINR}
\end{figure}

Fig. \ref{SINR} shows the angle estimation performance versus the scattering path power. The CRB is calculated given true user and scatter locations. It is observed that for low scattering path power and dense angle sampling, the estimation RMSE can approach the CRB. However, the effectiveness of scattering path elimination becomes limited for high scattering path interference, causing the RMSE to deviate from the CRB. Additionally, sparse angle sampling exacerbates the grid mismatch problem, further leading to accuracy degradation.

% \begin{figure}
% 	\centering
% 	\includegraphics[scale=0.48]{draw/simulation_new/fig6_1214.pdf}
% 	\caption{Angle estimation performance versus localization ranges for FF users.}
% 	\label{range}
% \end{figure}

% Fig. \ref{range} shows the angle estimation RMSE versus the localization range. We observe that as the localization range expands, the RMSE under the NF model increases, while the RMSE remains unchanged for the hybrid model. This is because as the localization range expands, for the NF model, the number of candidate locations also increases, while they remain the same for the FF and hybrid model. Also, note that for a sufficiently small estimation range, the NF model could outperform the hybrid model. This is due to that the NF model is more accurate and the FF model is the approximation of the NF model for the FF region. When the localization range is sufficiently small, the above issue of the NF model is mitigated, and the advantage of the NF model being more accurate exceeds the losses caused by the above problem. 

\subsection{Localization Accuracy and Complexity Trade-off}

This subsection examines the effects of hyperparameters on localization accuracy and algorithm complexity. We focus on the sampling spacing, the number of cycles, and the number of RIS elements, as they are the most important factors for both according to the analysis in Sec. \ref{section:peran}.

% \begin{figure*}[t!]
% 	\centering
% 	\includegraphics[scale=0.405]{draw/simulation_new/comp.pdf}
% 	\caption{(a) Angle estimation RMSE versus the number of cycles. (b) CPU time of the proposed algorithm versus number of cycles. (c) Localization of the NF user for different sampling spacing. (d) CPU time versus sampling spacing. (e) Range estimation performance of the NF users versus RIS size.\protect\footnotemark[1] (f) CPU time of the proposed algorithm for different RIS sizes.}
% 	\label{fig:comp}
% \end{figure*}

\subsubsection{Number of Cycles}

\begin{figure*}
	\centering
	\includegraphics[scale=0.5]{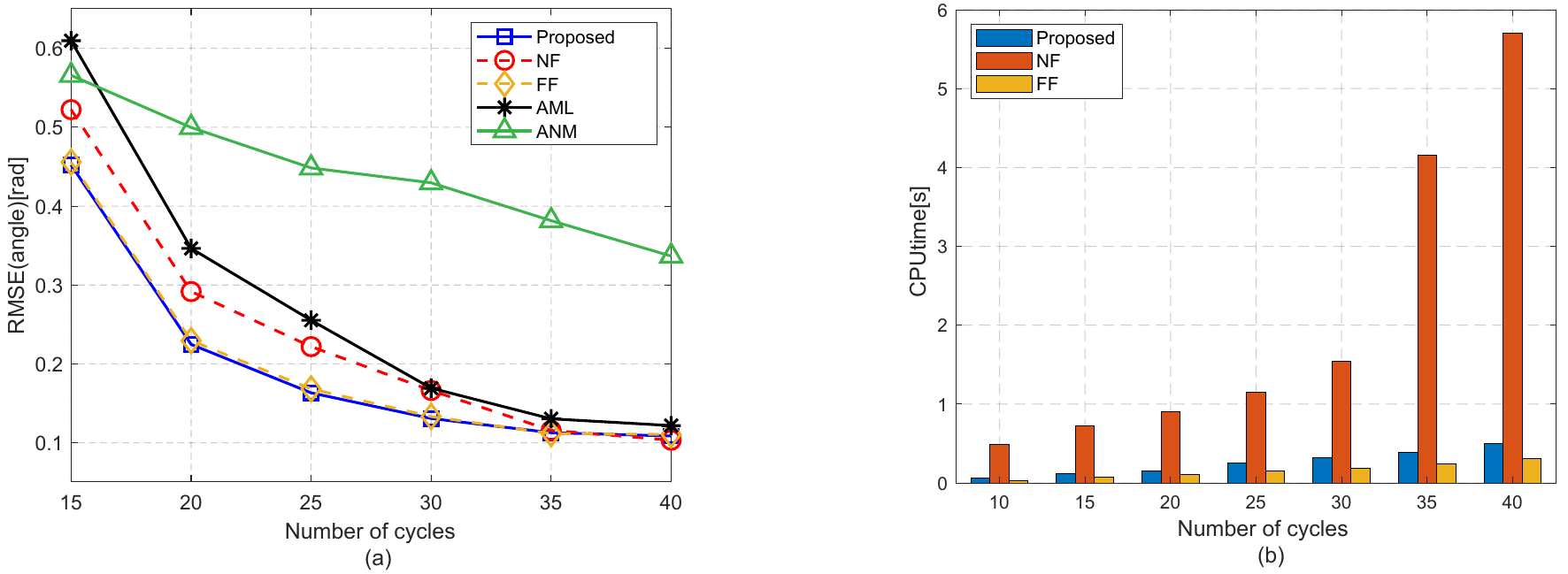}
	\caption{(a) Angle estimation RMSE versus the number of cycles. (b) CPU time of the proposed algorithm versus number of cycles.\protect\footnotemark[2]}
	\label{fig:cycles}
\end{figure*}

Fig. \ref{fig:cycles} (a) illustrates the angle estimation RMSE for different numbers of cycles. It shows that for the hybrid and FF model, the estimation performance converges after $C=30$ cycles, while for the NF model, the result remains fixed after $C= 35$ cycles, proving that the NF model is harder to converge. Fig. \ref{fig:cycles} (b) reveals the CPU time versus number of cycles for three models. The results show the proposed algorithm has the lowest complexity with FF model, followed by the hybrid model, and the highest complexity with NF model. This is because the FF model has the fewest candidate locations, while the NF model has the most. 

\begin{figure*}
	\centering
	\includegraphics[scale=0.5]{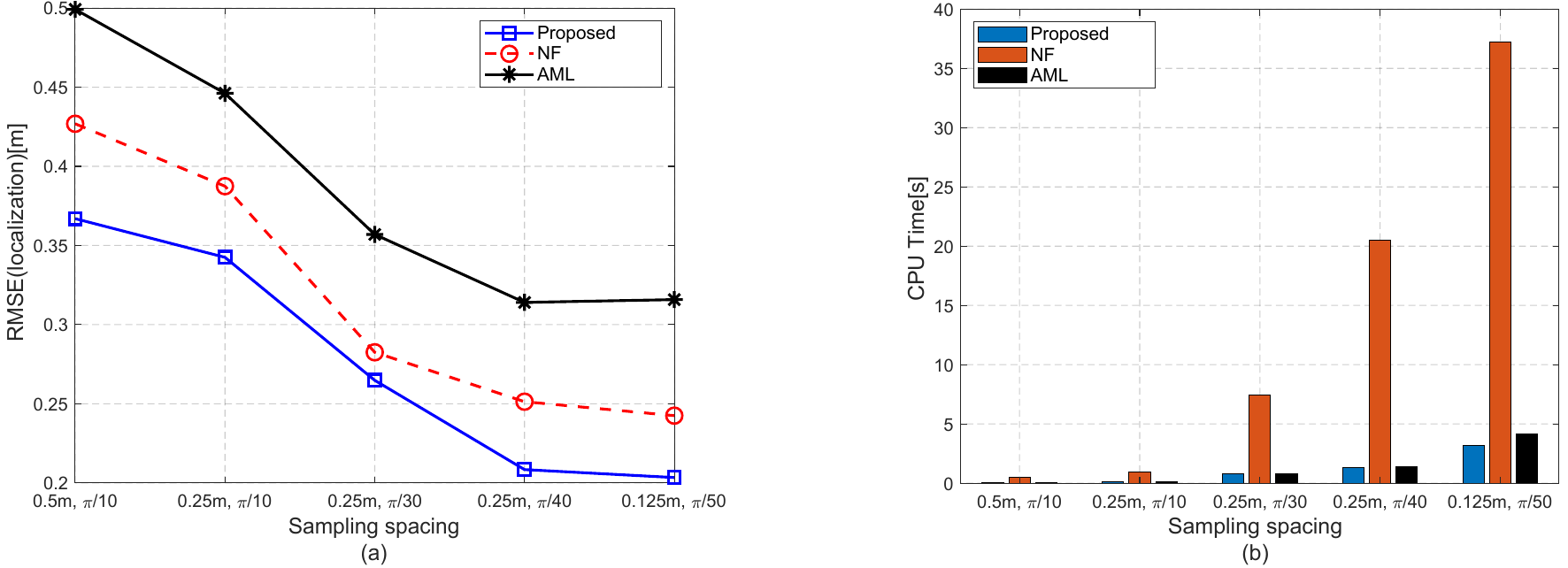}
	\caption{(a) Localization of the NF user for different sampling spacing. (b) CPU time versus sampling spacing.}
	\label{fig:sampling}
\end{figure*}

\subsubsection{Sampling Spacing}

Fig. \ref{fig:sampling} (a) shows the average localization RMSE for NF users of the proposed algorithm under different sampling spacing. The localization RMSE is derived by computing user positions from estimated angles and ranges, then comparing them to the ground truth. Fig. \ref{fig:sampling} (b) shows the CPU time of the proposed algorithm in three different models. It can be observed that as the sampling spacing reduces, the CPU time increases rapidly, while the estimation error first decreases and then remains fixed, which is consistent with previous analysis. Note that when the spacing is less than $(0.25\text{m}, \pi/40 \ \text{rad})$, the accuracy does not improve, but the complexity still increases.

\footnotetext[2]{Under the specified parameters at SNR = 5 dB, our method achieves an angle estimation error of $7^{\circ}$ and a range error of 0.2 m, with a computational latency of 0.15 s per update. This performance enables applications such as indoor navigation\cite{b91} and asset tracking\cite{b92}.}
%  and industrial robots localization\cite{b90}

\subsubsection{RIS Size}
\begin{figure*}
	\centering
	\includegraphics[scale=0.5]{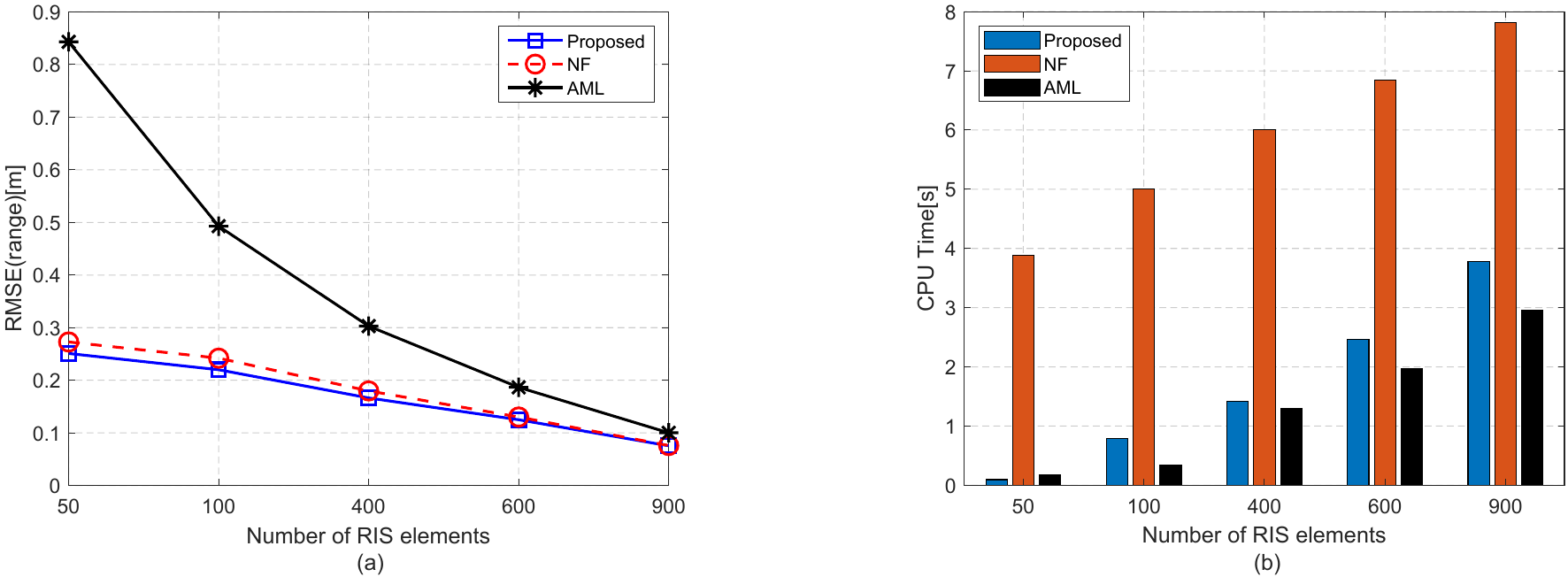}
	\caption{(a) Range estimation performance of the NF users versus RIS size. (b) CPU time of the proposed algorithm for different RIS sizes.}
	\label{fig:RISelement}
\end{figure*}

We set the number of cycles as $C=50$ to ensure convergence. Fig. \ref{fig:RISelement} (a) and Fig. \ref{fig:RISelement} (b) show the range estimation RMSE and the CPU time of the proposed algorithm versus the number of RIS elements. As the number of RIS elements increases, the range estimation RMSE decreases, and CPU time increases. This is because the complexities of both the localization and optimization algorithms are positively correlated with the RIS element number. Also, a centimeter-level accuracy is attainable for a RIS with $900$ elements.

% \protect\footnotemark[1] 
% \footnotetext[1]{For 100-element ($10 \times 10$) RIS, the NF range is approximately 0.9m. For 900-element ($30 \times 30$) RIS, the NF range is approximately 8m.}

\section{Conclusion}
\label{sec:con}
In this paper, a RIS-enabled multi-user hybrid NF and FF localization method has been developed. We considered a multi-user hybrid NF and FF localization scenario and introduced a localization protocol. A localization algorithm is designed for location estimation. The CRB of the estimated location parameters was analyzed and a CCM-based optimization method is proposed to optimize RIS phase shifts. We analyzed the complexity and the localization performance of the proposed method. Simulation results have shown that: 1) the proposed method outperforms the compared algorithm and can achieve sub-meter accuracy; 2) the hybrid NF and FF model has the advantages of high accuracy and low algorithm complexity over pure NF and FF models; 3) by adjusting the hyperparameters, the trade-off between localization accuracy and complexity can be achieved.

\begin{appendices}

\section{Proof of Proposition \ref{prop:NFCRB}}
\label{appen:NFCRB}
As mentioned in Section~\ref{local:formu}, all previously received signals are used to estimate the locations, so we need to consider received signals in all previous $c$ cycles, as modeled in (\ref{rsm}), to obtain the expressions of CRBs. We first derive the expression of the Fisher information matrix (FIM) related to the unknown location parameters according to the received signals, denoted by $\bm{J}_{NF}$. The CRB, which gives the localization performance limit, is then derived from the FIM. 

% Let $\hat{\bm{p}}$ denotes an unbiased estimate of a NF location $\bm{p}$, the MSE matrix of $\hat{\bm{p}}$ satisfies the following inequality\cite{b6}
% \begin{align}
% 	\mathbb{E}\{ (\hat{\bm{p}} - \bm{p})(\hat{\bm{p}} - \bm{p})^H\} \succeq \bm{J}_{NF}.
% \end{align}
As proved in \cite{b74}, FIM is given by\cite{b6}
\begin{align}
	[\bm{J}_{NF}]_{i,j}=E \left\{ -\frac{\partial^2 \ln f(\bm{y}_k;\bm{p}_k,\bm{Q})}{\partial p_i \partial p_j} \right\},
\end{align}
where $f(\bm{y}_k;\bm{p}_k,\bm{Q})$ is the likelihood function of the random vector $\bm{y}_k$ conditioned on $\bm{p}_k$ and $\bm{Q}$. Here we assume the transmitted signals are deterministic. Therefore the $i$-th received signal $y^{(i)}_k$ can be assumed to be independent of each other and follow complex Gaussian distribution, i.e., $f(y^{(i)}_k;\bm{p}_k,\bm{Q}) \sim \mathcal{CN}(\mu^{(i)},\sigma^2)$, where $\mu^{(i)}= (\bm{\beta}^{(i)})^T \bm{h}_ks$ is the noise-free received signal. Therefore, we have $f(\bm{y}_k)= \prod_{i=1}^{c+1}f(y^{(i)}_k)$.
% \begin{align}
% 	f(\bm{y}_k)= \prod_{i=1}^{c+1}f(y^{(i)}_k).
% \end{align}
Then, in the NF region, the FIM $\bm{J}_{NF}$ can be rewritten as
\begin{align}
	[\bm{J}_{NF}]_{i,j} = \dfrac{2}{\sigma^2} \sum_{m=1}^{c+1}Re \left \{ \dfrac{\partial (\mu^{(m)})^H}{\partial p_i} \dfrac{\partial \mu^{(m)}}{\partial p_j} \right \},
	\label{jp}
\end{align}
where $\frac{\partial \mu^{(m)}}{\partial p_i}=(\bm{\beta}^{(m)})^T \frac{\partial \bm{h}^{NF}_k}{\partial p_i}s$. The $n$-th entry of $\frac{\partial \bm{h}^{NF}_k}{\partial p_i}$ can be expressed as
\begin{align}
	\left[ \dfrac{\partial \bm{h}^{NF}_k}{\partial p_i}\right]_{n} =\alpha_k \left( -j\dfrac{2\pi}{\lambda} \right) \exp\left(-j\dfrac{2\pi}{\lambda}d_{kn}^t \right) \dfrac{\partial d_{kn}^t}{\partial p_i}.
\end{align}

Next, we take the inverse of $\bm{J}_{NF}$ and obtain the CRBs. Hence, the proof of proposition \ref{prop:NFCRB} is accomplished.

\section{Proof of Proposition \ref{prop:FFCRB}}
\label{appen:FFCRB}
Similar to the proof of proposition 2, the FIM $\bm{J}_{FF}$ in the FF region can be written as\cite{b6}
\begin{equation}
    [\bm{J}_{FF}]_{i,j} = \dfrac{2}{\sigma^2} \sum_{i=1}^{c+1}Re \left \{ \dfrac{\partial (\mu^{(i)})^H}{\partial p_i} \dfrac{\partial \mu^{(i)}}{\partial p_j} \right \},
\end{equation}
where $\frac{\partial \mu^{(i)}}{\partial p_i}=(\bm{\beta}^{(i)})^T \frac{\partial \bm{h}^{FF}_k}{\partial p_i}s$. In the FF region, we have
\begin{flalign}
    \!\!&\left[ \dfrac{\partial \bm{h}^{FF}_k}{\partial \theta_k}\right]_{n} \!\!\! \! =  \! h_{k0n}^t \! \left( \! -j \dfrac{2\pi}{\lambda} \! \right) \! ( \!-y_n \! \cos\theta_k \! \sin\phi_k \! + \! z_n \! \sin\theta_k), \!\!\! \\
    \!\!&\left[ \dfrac{\partial \bm{h}^{FF}_k}{\partial \phi_k}\right]_{n} \!\!\!\! =  \! h_{k0n}^t  \! \left( -j \dfrac{2\pi}{\lambda} \right)  (-y_n \sin\theta_k \cos\phi_k),
\end{flalign}
where $h_{k0n}^t$ is the $n$-th element in the direct user-RIS path.

Next, we take the inverse of $\bm{J}_{FF}$ and obtain the CRBs. Hence, the proof of proposition \ref{prop:FFCRB} is accomplished.

\section{The Differentials of CRB}
\label{appen:dCRB}
% We denote the Fisher information matrix $\bm{J}$ in the NF case as
% \begin{equation}
% \bm{J}_{NF}
% =
% \left[
% \begin{array}{ccc}
%     J_{11} & J_{12} & J_{13} \\
%     J_{21} & J_{22} & J_{23} \\
%     J_{31} & J_{32} & J_{33}
% \end{array}
% \right],
% \end{equation}
% and the Fisher information matrix in the FF case as
% \begin{equation}
% \bm{J}_{FF}
% =
% \left[
% \begin{array}{ccc}
%     J_{11} & J_{12} \\
%     J_{21} & J_{22} \\
% \end{array}
% \right].
% \end{equation}

We denote the element in the $i$-th row, $j$-th column of $\bm{J}$ as $J_{ij}$. In order to find the differentials of CRBs, we need to find the differentials of $J_{ij}$.
For simplicity, here we use $\bm{\beta}$ to substitute $\bm{\beta}^{(k+1)}$. $J_{ij}$ can be considered as a function of two independent complexed-valued variables $\bm{\beta}$ and $\bm{\beta ^{\ast}}$. The complex differentials of $J_{ij}$ can be expressed as
\begin{align}
		dJ_{ij} =  \left( \dfrac{\partial J_{ij}}{\partial \bm{\beta}} \right) d \bm{\beta} +  \left( \dfrac{\partial J_{ij}}{\partial \bm{\beta ^{\ast}}} \right) d \bm{\beta}^{\ast}.
\end{align}
The complex differentials of $J_{ij}$ are given as
\begin{align}
\label{equation:dj}
	dJ_{ij}
	=& \dfrac{1}{\sigma^2} \left[ d \left( s^H \left(\dfrac{\partial \bm{h}}{\partial p_i} \right)^H \bm{\beta}^{\ast} \bm{\beta}^T \dfrac{\partial \bm{h}}{\partial p_j}  s \right) \right. \nonumber \\
	& + \left. d \left ( s^T \left(\dfrac{\partial \bm{h}}{\partial p_i} \right)^T \bm{\beta} \bm{\beta}^H \left( \dfrac{\partial \bm{h}}{\partial p_j}\right)^{\ast}  s^{\ast} \right) \right ].
\end{align}
By substitude $\frac{\partial \bm{\beta}^{\ast} \bm{\beta}^T}{\partial \bm{\beta}_n^{\ast}}$ and $\frac{\partial \bm{\beta} \bm{\beta}^H}{\partial \bm{\beta}_n^{\ast}}$ into (\ref{equation:dj}), we can obtain 
\begin{align}
	\dfrac{\partial J_{ij}}{\partial \beta_n^{\ast}}  = \dfrac{|s|^2}{\sigma^2} \left[ \left( \dfrac{\partial h_n}{\partial p_i} \right)^{\ast} \sum_{m=1}^{N} \dfrac{\partial h_m}{\partial p_j}  \beta_m + \right. \nonumber \\
    \left. \left(  \dfrac{\partial h_n}{\partial p_j} \right)^{\ast}  
    \sum_{m=1}^{N}\dfrac{\partial h_m}{\partial p_i} \beta_m  \right].
\end{align}

The derivatives of CRBs with respect to $\bm{\beta}^{\ast}$ are given by
\begin{align}
    \ \ \dfrac{\partial D_i}{\partial \bm{\beta}^{*}} &= \left(\dfrac{\partial \bm{J}^{-1}}{\partial \bm{\beta}^{*}}\right)_{i,i} = \left( - \bm{J}^{-1} \dfrac{\partial \bm{J}}{\partial \bm{\beta}^{*}}\bm{J}^{-1} \right)_{i,i}.
\end{align}

\end{appendices}

\bibliographystyle{IEEEtran}
\bibliography{IEEEabrv,ref}

\end{document}